\documentclass[12pt]{article}

\usepackage{latexsym,amsmath,amsfonts,amssymb}
\usepackage[latin1]{inputenc}
\usepackage[american]{babel}
\usepackage[dvips]{graphicx}
\usepackage{bbm}
\usepackage{bm}
\usepackage{subfigure}
\usepackage{url}
\usepackage{multirow}

\renewenvironment{thebibliography}[1]{%
\begin{oldthebibliography}{#1}%
\setlength\itemsep{-1mm}%
}%
{%
\end{oldthebibliography}%
}

\newcommand{\re}{\,\mathbb{R}\mbox{e}\,}
\newcommand{\im}{\,\mathbb{I}\mbox{m}\,}

\newcommand{\ud}[2]{^{#1}_{\phantom{#1}#2}}

\newcommand{\eg}{\textit{e.g.}}

\numberwithin{equation}{section}
\numberwithin{equation}{section}

\newcommand{\nn}{\nonumber}

\newcommand{\be}{\begin{equation}} \newcommand{\ee}{\end{equation}}
\newcommand{\bea}{\begin{equation} \begin{aligned}} \newcommand{\eea}{\end{aligned} \end{equation}}

\newcommand{\cF}{\mathcal{F}}

\newcommand{\cJ}{\mathcal{J}}

\newcommand{\cN}{\mathcal{N}}

\newcommand{\cR}{\mathcal{R}}

\newcommand{\bZ}{\mathbb{Z}}

\newcommand{\unit}{\mathbbm{1}}
\newcommand{\ts}{\tilde{s}}

\newcommand{\Gh}{\mathrm{\Gamma}_h}

\DeclareMathOperator{\Tr}{Tr}

\DeclareMathOperator{\sign}{sign}
\DeclareMathOperator{\rank}{rank}
\DeclareMathOperator{\diag}{diag}

\newcommand{\eps}{\epsilon}

\begin{document}

\makeatletter \@addtoreset{equation}{section} \makeatother
\renewcommand{\theequation}{\thesection.\arabic{equation}}
\pagestyle{empty}
\rightline{PU-2391}
\rightline{WIS/07/11-AUG-DPPA}
\rightline{TAUP-2931/11}
\vspace{0.8cm}
\begin{center}
{\LARGE{\bf Comments on 3d Seiberg-like dualities
 \\[10mm]}} {\large{Francesco Benini$^{1}$,
 Cyril Closset$^{2}$,  Stefano Cremonesi$^{3}$ \\[5mm]}}
{\small{{}$^1$ Department of Physics, Princeton University \\
\vspace*{-2pt} Princeton, NJ 08544, USA\\

\medskip
{}$^2$ Department of Particle Physics and Astrophysics \\
Weizmann Institute of Science, Rehovot 76100, Israel \\

\medskip
{}$^3$
Raymond and Beverly Sackler School of Physics and Astronomy\\
\vspace*{-2pt} Tel-Aviv University, Ramat-Aviv 69978, Israel}}

\medskip

\medskip

\medskip

\medskip

\medskip

{\bf Abstract}
\vskip 20pt
\begin{minipage}[h]{16.0cm}
We study Seiberg-like dualities in three dimensional $\cN=2$ supersymmetric theories, emphasizing Chern-Simons terms for the global symmetry group, which affect contact terms in two-point functions of global currents and are essential to the duality map.
We introduce new Seiberg-like dualities for Yang-Mills-Chern-Simons theories with unitary gauge groups with arbitrary numbers of matter fields in the fundamental and antifundamental representations. These dualities are derived from Aharony duality by real mass deformations. They allow to initiate the systematic study of Seiberg-like dualities in Chern-Simons quivers. We also comment on known Seiberg-like dualities for symplectic and orthogonal gauge groups and extend the latter to the Yang-Mills case. We check our proposals by showing that the localized partition functions on the squashed $S^3$ match between dual descriptions.
\end{minipage}
\end{center}

\newpage

\setcounter{page}{1} \pagestyle{plain}
\renewcommand{\thefootnote}{\arabic{footnote}} \setcounter{footnote}{0}

{
\tableofcontents}

\vspace*{1cm}


\section{Introduction}
Dualities in supersymmetric gauge theories are powerful tools providing a better understanding of the strongly coupled regime. In this paper we study some dualities in Yang-Mills (YM) and Yang-Mills-Chern-Simons (YM-CS) theories in three dimensions preserving 4 supercharges ($\cN=2$). In three dimensions the Yang-Mills coupling $g^2$ is dimensionful, making the theory super-renormalizable but often strongly coupled in the infrared (IR). The theories we consider are believed to flow to interacting fixed points in the IR. In the case where a Chern-Simons term is present one can also define the theory without Yang-Mills coupling, so that the Lagrangian is classically marginal; such a theory is also exactly superconformal in the sense of \cite{Gaiotto:2007qi}, albeit strongly coupled unless the CS level $k$ or the number of flavors is large, and it can be thought of as the IR limit of a YM-CS theory.

The type of duality we want to consider is an exact equivalence between the IR fixed points of two distinct YM-CS theories. Such an ``IR duality'' is reminiscent of four-dimensional $\cN=1$ Seiberg duality in the conformal window \cite{Seiberg:1994pq}.
There are roughly two kinds of IR dualities which are known to exist in 3d $\cN=2$ gauge theories. The first kind is known as mirror symmetry \cite{Intriligator:1996ex, deBoer:1997ka}: its hallmark is that it exchanges the role of fundamental fields and monopole operators \cite{Aharony:1997bx, Borokhov:2002cg}. The second kind of duality, more akin to Seiberg duality \cite{Aharony:1997bx, deBoer:1997kr, Karch:1997ux, Aharony:1997gp, Niarchos:2009aa}, has recently attracted renewed attention \cite{Aharony:2008gk, Giveon:2008zn, Kapustin:2010mh, Willett:2011gp, Kapustin:2011gh, Bashkirov:2011vy, Dolan:2011rp, Berenstein:2011dr, Morita:2011cs}.
 In this paper we will focus on such ``Seiberg-like'' dualities.

We stress that the Chern-Simons terms for global symmetries, which we call ``global CS terms'', can change under duality (section \ref{sec: global CS terms}).
We consider (section \ref{sec: SD for chiral SQCD--general}) novel Seiberg-like dualities for YM-CS theories with unitary gauge groups and generic number of fundamental and antifundamental matter representations. They can be obtained from dual pairs of theories considered by Aharony \cite{Aharony:1997gp} via RG flows triggered by real mass deformations.%
\footnote{Some of these flows to vectorlike theories were previously analyzed in \cite{ItamarShamir}.}
We are careful to specify the relative global CS terms that arise in these dualities. These terms become especially important when (part of) the global symmetries are gauged within a larger theory, \eg{} a quiver (section \ref{sec: quivers}). We support our claims (section \ref{sec: Un duality and S3 PF}) by showing that the partition functions $Z$ on the squashed 3-sphere $S^3_b$ \cite{Hama:2011ea} agree on the two sides of the dualities. By numerical computation (section \ref{sec: Z-min}) we can check that $|Z|$ increases along the RG flow in some simple cases, giving support to the conjectured ``F-theorem'' \cite{Jafferis:2011zi}. We further comment (sections \ref{sec: Usp} and \ref{sec: O}) on previously known dualities for the case of symplectic and orthogonal gauge groups and extend the orthogonal dualities of \cite{Kapustin:2011gh} to Yang-Mills theories without CS interactions.%
\footnote{The latter duality is also proposed in \cite{Aharony:2011ci}, which appeared soon after the first version of this paper.}
These results can be used to check the duality between $\cN=5$ $Usp(2N_{Usp})_{k}\times O(N_O)_{-2k}$ theories put forward in ABJ \cite{Aharony:2008gk} from the partition function.

\section{Global Chern-Simons terms}
\label{sec: global CS terms}

Let us consider YM-CS theories with a semi-simple global symmetry group $G$. We assume that there are no accidental symmetries in the IR. Let $J_I$, $I=1, \cdots, \rank(G)$ be the conserved currents in the Cartan subalgebra of $G$. The two-point function of conserved currents in three dimension can contain a conformally invariant contact term \cite{Witten:2003ya}
\be\label{contact terms}
\langle J_I^{\mu}(x)\, J_J^{\nu}(y) \rangle \, =\,-\frac{w_{IJ}}{4\pi} \epsilon^{\mu\nu\rho} \partial_{\rho}\, \delta^3(x-y)  +\cdots\, .
\ee
The coefficients $w_{IJ}$ are part of the definition of the conformal theory in the IR.

In the UV where we have a Lagrangian description, the contact terms (\ref{contact terms}) correspond to Chern-Simons interactions for the background gauge fields coupling to the conserved currents,
\be\label{S background for G 01}
S_{\text{background}} = \sum_{I, J} \frac{k_{IJ}}{4\pi}\int_{M_3} A_I \wedge dA_J\, ,
\ee
where $M_3$ is the manifold on which the theory is defined. We have $w^{UV}_{IJ}=k_{IJ}$.
The formulas for the non-Abelian symmetries are similar.
For a pair of dual theories with the same IR physics, it might happen that there is a relative global Chern-Simons term so that the IR contact terms $w_{IJ}$  in (\ref{contact terms}) agree. This was noticed previously in the context of mirror symmetry \cite{Kapustin:1999ha} and we find this also to be the case in many Seiberg-like dualities.

Let us further comment on the $\cN=2$ supersymmetric completion of such interactions. Whenever $J_I$ is the current of a non-R Abelian symmetry, $J_I^{\mu}$ is simply the $\theta\bar{\theta}$ component of a linear superfield $\cJ_I$ (with $D^2\cJ_I= \bar{D}^2\cJ_I=0$), it couples to a background vector superfield $V_I$,  and the $\cN=2$ completion of (\ref{S background for G 01}) is
\be\label{S background for G 02}
 \sum_{I, J} \frac{k_{IJ}}{4\pi}  \int d^3x \int d^4\theta  \, V_I \Sigma_J\, ,
\ee
where $\Sigma_J = \epsilon^{\alpha\beta} \bar{D}_\alpha D_\beta V_J$ is the linear multiplet containing the field strength of $V_J$.
On the other hand, the supersymmetric gauging of the R-symmetry is more subtle; see \cite{Komargodski:2010rb, Hama:2011ea, Festuccia:2011ws} for recent discussions. The R-symmetry current $J_R^{\mu}$ is the lowest component of a R-multiplet $\cR_{\mu}$ including the supercurrent and the stress-energy tensor.
It can be coupled minimally to supergravity:  there exists a supergravity multiplet%
\footnote{It is the dimensional reduction from 4 to 3 dimensions of the new-minimal SUGRA multiplet of \cite{Sohnius:1981tp}. }
 whose highest component is a gauge field $A_{\mu}^{(R)}$ for local R-symmetry transformations (it is an  auxiliary field in the off-shell SUGRA multiplet).
Therefore there exists an $\cN=2$ completion of the CS term
\be
\frac{k_{R R}}{4\pi} \int A^{(R)}\wedge d A^{(R)}\ ,
\ee
which involves supersymmetry preserving background fields in the SUGRA multiplet, in the spirit of \cite{Hama:2011ea, Festuccia:2011ws}. A full description of this is left for the future.

\section{Dualities for $U(n)$ theories with ``chiral'' matter}
\label{sec: SD for chiral SQCD--general}

\begin{figure}[t]
\begin{center}
\subfigure[\small Electric]{
\includegraphics[width=5cm]{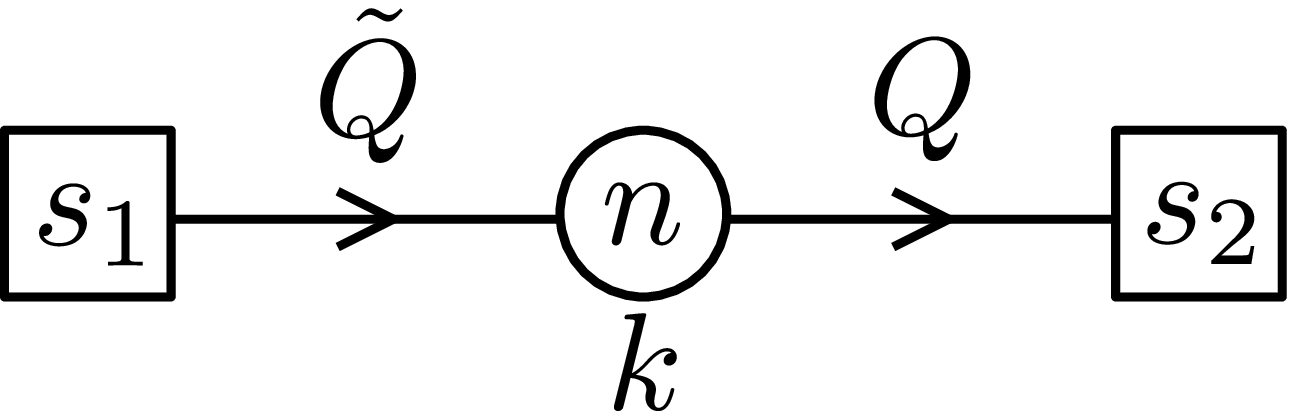}
}
\qquad \qquad
\subfigure[\small Magnetic]{
\includegraphics[width=5cm]{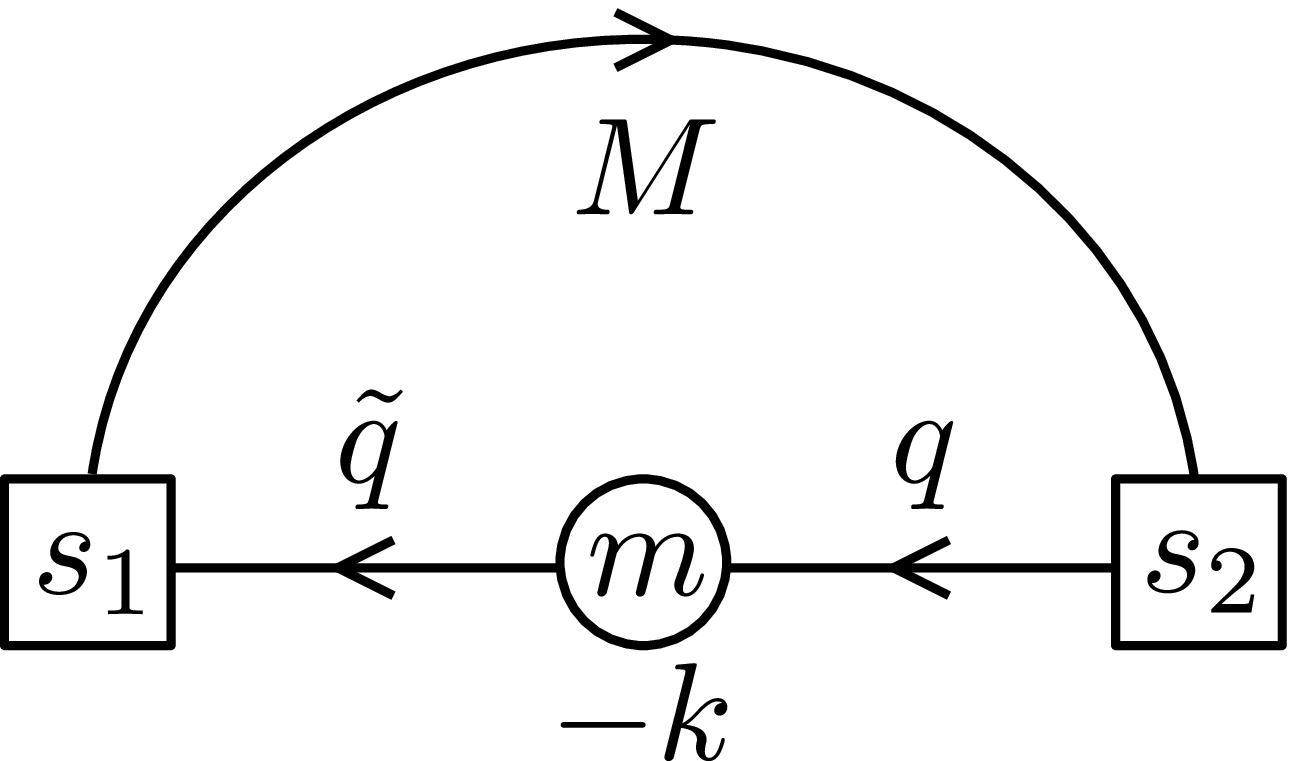}}
\end{center}
\caption{\small (a) Quiver diagram of the $U(n)_k$ electric theory. (b) Quiver diagram of the magnetic $U(m)_{-k}$ theory, where $m$ is defined in (\ref{m for various cases}). For special values of the parameters, there might be extra singlets $T$ and/or $\tilde{T}$ in addition to the mesons $M$, as explained in the text.
\label{fig: quiver}}
\end{figure}

Let us consider the following ``SQCD-like" theory in three dimensions. We take a $\cN=2$ supersymmetric gauge theory with gauge group $U(n)$, coupled to $s_1$ matter superfields $\tilde{Q}^a$ in the antifundamental representation $\bm{\overline{n}}$ and $s_2$ matter superfields $Q_b$ in the fundamental representation $\bm{n}$; in short we denote this matter content $(s_1, s_2)$, and the quiver diagram is in figure \ref{fig: quiver} (left). In addition to the super-Yang-Mills couplings for the $U(n)$ vector multiplet, we allow for $\cN=2$ Chern-Simons interactions, with Chern-Simons level $k$.
Due to the absence of chiral anomalies in three dimensions, we can take $s_1 \neq s_2$, a situation we loosely call ``chiral''.
On the other hand, cancelation of a $\bZ_2$ anomaly \cite{Redlich:1983kn, Redlich:1983dv} requires that
\be
k + \frac12(s_1+s_2) \,\in \bZ\, .
\ee
This theory is expected to flow to an interacting fixed point in the IR, generally strongly coupled.
Dualities can be very useful to obtain more clues about the IR dynamics, for instance about the chiral ring of superconformal primaries.

\subsection{Aharony's duality}
\label{sec: aharony dual}

The case of a $U(n)$ gauge group with $k=0$ and matter content $(n_f, n_f)$ has been studied in detail over the years \cite{Aharony:1997bx, deBoer:1997kr, Aharony:1997gp, Dorey:1998kq}.
Classically this three-dimensional SQCD theory has both a Coulomb and a Higgs branch. The Higgs branch corresponds to VEVs for the gauge-invariant mesons $M= \tilde{Q}Q$. The Coulomb branch corresponds to VEVs for the adjoint scalar $\sigma$ in the vector multiplet. Since a generic VEV for $\sigma$ breaks the gauge group to its maximal torus and it gives mass to all quarks, the low energy theory along the Coulomb branch consists of free photons (and their superpartners) for the unbroken $U(1)^n$ gauge group. The free vector multiplets can be dualized into chiral superfields $\Phi_i = \frac{2\pi }{e^2} \sigma_i + i \varphi_i$, where $\varphi_i$ are dual photons (of period $2\pi$), and the classical Coulomb branch is parameterized by monopole operators $Y_i \sim \exp{\Phi_i}$. However instanton corrections lift most of the Coulomb branch \cite{Aharony:1997bx, deBoer:1997kr, Aharony:1997gp}. The only remaining monopole operators are the ones with flux $H=(\pm 1, 0, \cdots, 0)$ in the Cartan of the gauge group, which we denote as $T$ and $\tilde{T}$. They have charge $\pm 1$ under the topological symmetry with current $\Tr (\ast F)$. At the quantum level, for $n_f < n-1$ a dynamically generated superpotential lifts all SUSY vacua. For $n_f = n-1$ the singular classical moduli space is deformed quantum-mechanically. We will focus on the case $n_f \geq n$, for which there is an effective superpotential on the moduli space,
\be
\label{Weff for Aharony theory}
W_{\text{eff}} = (n_f-n+1) (T\tilde{T} \det M)^{\frac{1}{n_f-n+1}} \; .
\ee
In the case $n_f=n$ the IR physics is described by a sigma model for the chiral superfields $M$, $T$ and $\tilde{T}$.
For $n_f > n$ the effective superpotential (\ref{Weff for Aharony theory}) correctly describes the moduli space%
\footnote{For $\tilde{T}=T=0$ we have $\rank(M)\leq  n$; if either $T$ or $\tilde{T}$ is non-zero we have $\rank(M)\leq  n-1$, while if both $T$ and $\tilde{T}$ are non-zero then $\rank(M)\leq  n-2$. In particular the semi-classical result that $M= \tilde{Q}Q$ has at most rank $n < n_f$ is reproduced.}
except for the origin $M=T=\tilde{T}=0$ where it is singular. In 4d $\cN=1$ SQCD it is well known that the same problem is cured by the introduction of a dual gauge group coupled to dual quarks which can be weakly coupled (below the 4d conformal window) \cite{Seiberg:1994pq}. In three dimensions the gauge coupling is never IR free, but a situation similar to the 4d conformal window, with two different theories flowing to the same IR fixed point, can occur. Such a Seiberg-like duality was proposed by Aharony a while ago \cite{Aharony:1997gp}.

Let us call the $\cN=2$ supersymmetric $U(n)$ gauge theory with matter content $(n_f, n_f)$, denoted  $(\tilde{Q}^a, Q_b)$, the ``electric'' theory. The ``magnetic'' theory of \cite{Aharony:1997gp} is a $U(n_f-n)$ gauge theory with matter content $(n_f, n_f)$, denoted $(q^a, \tilde{q}_b)$, together with $n_f^2$ singlets $M^a_b$, two singlets $T$ and $\tilde{T}$ and a superpotential
\be\label{W for magnetic dual of Ofer}
W= \tilde{q} M q  + tT + \tilde{t}\tilde{T}\, ,
\ee
where $t$ and $\tilde{t}$ are the monopole operators in the $U(n_f-n)$ gauge group with Abelian flux $+1$ and $-1$ respectively.
The description of the magnetic theory is to be taken with a grain of salt, because of the appearance of monopole operators in $W$ \cite{Aharony:1997gp}. The significance of (\ref{W for magnetic dual of Ofer}) is that we have the relations
\be\label{F term rel in dual of ofer}
q\tilde{q}=0 \;,\qquad t=0 \;,\qquad \tilde{t}=0 \;,
\ee
in the chiral ring of the magnetic theory.
We also have
\be\label{F term rel in dual of ofer Bis}
\tilde{q}M = 0 \;,\qquad Mq=0 \;,
\ee
which tell us that for generic $q$, $\tilde{q}$ the meson matrix $M$ has at least $n_f-n$ zero eigenvalues, so that $\rank{M}\leq n$ like in the electric theory.
The representations under gauge and global symmetries of the fields in both theories are:
\be
\label{charges for aharony duals}
\begin{array}{c|cc|ccccc}
    &  U(n)& U(n_f-n)& SU(n_f) & SU(n_f)  & U(1)_A &  U(1)_M & U(1)_R  \\
\hline
\tilde{Q}   & \bm{\overline{n}} &\bm{1}&    \bm{n_f} & \bm{1} & 1   & 0   &r_Q \\
Q        & \bm{n} & \bm{1} & \bm{1} & \bm{\overline{n_f}}& 1   & 0   &r_Q \\
\hline
\tilde{q}   & \bm{1} &\bm{n_f-n}&    \bm{\overline{n_f}} & \bm{1} & -1   & 0   &1-r_Q \\
q        & \bm{1} & \bm{\overline{n_f-n}} & \bm{1} & \bm{n_f}     & -1   & 0   &1-r_Q \\
M        & \bm{1} & \bm{1} & \bm{n_f} & \bm{\overline{n_f}}& 1   & 0   &2 r_Q \\
T      & \bm{1} & \bm{1} & \bm{1} & \bm{1} & -n_f   & 1   & n_f(1-r_Q) -n +1\\
\tilde{T}      & \bm{1} & \bm{1} & \bm{1} & \bm{1} & -n_f   & -1   & n_f(1-r_Q) -n +1\\
\hline
t              & \bm{1} & \bm{1} & \bm{1} & \bm{1} & n_f   & -1   & -n_f(1-r_Q) +n +1\\
\tilde{t}      & \bm{1} & \bm{1} & \bm{1} & \bm{1} & n_f   & 1   & -n_f(1-r_Q) +n +1
\end{array}
\ee
The duality identifies $M$ with the mesons $\tilde{Q}Q$, and the singlets $T$, $\tilde{T}$ with the gauge-invariant monopole operators of the electric theory. In the magnetic theory we also have monopole operators $t$ and $\tilde{t}$ which survive instanton corrections of the $U(n_f-n)$ gauge group.%
\footnote{Note that the topological charge $M= M_e$ of the electric theory is related to the topological charge $M_m$ of the magnetic theory by $M_e = -M_m$.}
However they vanish in the chiral ring, due to (\ref{F term rel in dual of ofer}).

This duality has received renewed attention recently: it has been checked at the level of the partition function in \cite{Willett:2011gp} and the matching of the superconformal index between the dual theories was discussed in \cite{Bashkirov:2011vy}.

\subsection{Chiral SQCD with Chern-Simons levels}
\label{sec:Chiral SQCD with Chern-Simons levels}

In the more general case $s_1 \neq s_2$, the Higgs branch of the $U(n)$ theory is still parametrized by mesons $M= \tilde{Q}Q$, but the Coulomb branch is generically lifted. Indeed, for a generic VEV $\sigma = \diag{(x_1, \cdots, x_n)}$, integrating out the massive quarks generates Chern-Simons terms for the vector multiplets in $U(1)^n$. For $U(1)_{i} \subset U(1)^n$, we have
\be
k_{i}^{\text{eff}} = k \,  -\frac12(s_1-s_2)\sign(x_{i})\, ,
\ee
where $k$ is the (possibly vanishing) bare CS level of the $U(n)$ gauge group.  A non-zero Chern-Simons level gives a so-called topological mass $m_T= g^2 k/2\pi$ \cite{Deser:1981wh} to the Abelian vector multiplets, lifting the Coulomb branch at one-loop, even before instanton corrections are taken into account. Let us define the effective CS levels
\be\label{k+-}
k_{\pm} \equiv k \,   \pm \frac12(s_2-s_1) \, .
\ee
There are four interesting cases according to the signs of $k_+$ and $k_-$, which we denote by
\bea\label{The chiral electric theories}
&[\bm{p},\bm{q}]a&= & \;[-k_+, -k_-]a\, , & \qquad \quad &  [\bm{p},\bm{q}]b &= &\; [k_+, k_-]b\, , & \\
&[\bm{p},\bm{q}]^*a&= &\; [-k_+, k_-]^*a\, ,  & \qquad \quad &  [\bm{p},\bm{q}]^*b &= &\; [k_+, -k_-]^*b\, . & \\
\eea
The labels $[\bm{p}, \bm{q}]$ are by definition non-negative integers. In fact there are only two inequivalent cases, say $[\bm{p},\bm{q}]a$ and $[\bm{p},\bm{q}]^*a$, because the $b$-cases are related to them by a CP transformation. We will therefore consider the $a$-cases in the following. The limiting cases between the $a$ and $^*a$ theories, such as $[\bm{p}, \bm{0}]a$, are quite interesting: one of the two effective CS levels vanishes and half of the Coulomb branch remains unlifted.

The global symmetry group of our theories is $SU(s_1)\times SU(s_2)\times U(1)_A \times U(1)_M \times U(1)_R$, unless $s_1s_2=0$ in which case $U(1)_A$ is not present. We have
\be
\label{charges of chiral electric theory}
\begin{array}{c|c|ccccc}
& U(n) & SU(s_1) & SU(s_2)   & U(1)_A & U(1)_{M} & U(1)_{R_0}  \\
\hline
\tilde{Q} & \bm{\overline{n}} & \bm{s_1} & \bm{1}  & 1& 0& 0 \\
Q &\bm{n} &\bm{1} & \bm{\overline{s_2}}  &  1&0& 0
\end{array}
\ee
The fundamental fields are neutral under the topological symmetry $U(1)_M$, but monopole operators are charged under it.
The $U(1)_{R_0}$ defined here is a convenient choice of UV R-charge, corresponding to $r_Q=0$ in (\ref{charges for aharony duals}). We will discuss the superconformal R-symmetry in more detail in section \ref{sec: Z-min}.

These theories either break supersymmetry spontaneously (in some cases due to a runaway), or they enjoy a Seiberg-like dual description.%
\footnote{The only exception is the case $k=0$, $s_1=s_2=n_f= n-1$, where the classical moduli space is deformed quantum-mechanically \cite{Aharony:1997bx}. \label{footnote}}
Whenever it exists, the ``magnetic'' theory is a $U(m)$ gauge theory at Chern-Simons level $-k$, with $s_1$ chiral superfields $\tilde{q}_a$ in the fundamental representation and $s_2$ chiral superfields $q^b$ in the antifundamental representation -- the quiver is in figure \ref{fig: quiver} (right). The rank $m$ of the dual gauge group is
\bea
\label{m for various cases}
& [\bm{p}, \bm{q}]a& \quad : \quad &m&=&\; \frac{s_1+s_2}{2} -k -n \, ,&\\
& [\bm{p}, \bm{q}]b& \quad : \quad&m&=&\; \frac{s_1+s_2}{2} +k -n \, ,&\\
& [\bm{p}, \bm{q}]^*a& \quad : \quad&m&=&\; s_1 -n  \, ,&\\
& [\bm{p}, \bm{q}]^*b& \quad : \quad&m&=&\; s_2-n \, .&\\
\eea
In addition, we have $s_1\times s_2$ gauge singlets $M^a_b$, dual to the mesons $\tilde{Q}^aQ_b$. In the case $[\bm{p},\bm{0}]a$ we also have an additional singlet $T$ dual to a gauge invariant monopole operator, and in the case $[\bm{0},\bm{0}]$ we have the duality of \cite{Aharony:1997gp}. The singlets are coupled to the magnetic sector through the superpotential
\be\label{W for magnetic dual, in general}
W_{\text{mag}}= \tilde{q} M q \,+ \,\delta_{\bm{p 0}}\, tT \,+\, \delta_{\bm{q0}}\, \tilde{T}\tilde{t} \;.
\ee
Note that the Giveon-Kutasov duality \cite{Giveon:2008zn} is the case $[\bm{p}, \bm{p}]a$. Remark that the dual of a theory of type $[\bm{p}, \bm{q}]a$ is a theory of type $[\bm{p}, \bm{q}]b$, with extra singlets coupled through (\ref{W for magnetic dual, in general}); similarly the dual of a theory of type $[\bm{p}, \bm{q}]^*a$ is a theory of type $[\bm{p}, \bm{q}]^*b$.

Supersymmetry breaking occurs when the parameter $m$ in (\ref{m for various cases}) is negative (recall however footnote \ref{footnote}). Consider a theory in any of the four cases (\ref{The chiral electric theories}), with matter content $(s_1, s_2)$ and $m>0$. If we give a complex mass to a pair $(\tilde{Q},Q)$ by a superpotential term $W=\mu \tilde{Q}^{s_1}Q_{s_2}$, we flow to a theory of the same type with matter content $(s_1-1, s_2-1)$. In the magnetic dual, this correspond to a linear term $\mu M^{s_1}_{s_2}$ in $W_{\text{mag}}$, which leads to a VEV for $q^{s_2} \tilde{q}_{s_1}$ and breaks the dual gauge group to $U(m-1)$, consistently with the proposed duality. On the other hand, if we start with a theory at $m=0$, the complex mass deformation leads to a superpotential $W_{\text{mag}}= \mu M^{s_1}_{s_2}$ in the dual theory, which breaks supersymmetry spontaneously.

It is interesting to consider generic real mass parameters for our theories. All the mass terms can be understood as background values for scalars in external vector multiplets which gauge the global symmetries. We have:
\bea\label{def real masses}
\tilde{m}_a &\qquad  :\quad & SU(s_1) & \qquad & a=1, \cdots, s_1\\
m_b &\qquad  :\quad & SU(s_2) & \qquad & b=1, \cdots, s_2 \\
m_A &\qquad  :\quad & U(1)_A & \qquad &  \\
\xi &\qquad  :\quad & U(1)_M & \qquad &
\eea
We take $\tilde{m}_a$ to correspond to the Cartan of $U(s_1)$, but we add the constraint $\sum_{a} \tilde{m}_a =0$ to get to $SU(s_1)$. Similarly $\sum_b m_b =0$ for $SU(s_2)$. The term $\xi$ is best thought of as a Fayet-Iliopoulos (FI) parameter for the diagonal $U(1)$ in the gauge group.
The quarks $\tilde{Q}$, $Q$ have effective real mass
\bea
\label{real masses for Q Qt in general}
M[\tilde{Q}^a_{i}] &=& -x_{i} + \tilde{m}_a + m_A\, , \\
M[Q^{i}_b] &=& x_{i} - m_b + m_A\, ,
\eea
where $x_{i}$ are the eigenvalues of $\sigma$. The duality we consider holds for generic massive deformations, provided that the magnetic theory has some definite Chern-Simons terms for the flavor group, which we will derive.

The numerology (\ref{m for various cases}) for the dual gauge group can also be found from a Type IIB brane construction \cite{Cremonesi:2010ae}, where the statement that SUSY is broken for $m <0$ follows from the s-rule of \cite{Hanany:1996ie}.
Here we prefer to keep the discussion in purely field theory terms.


\subsection{Dual of the $[\bm{p}, \bm{0}]a$ theory}
\label{subsec: dual of [p,0]a theory}

All Seiberg-like dualities we consider for $U(n)$ gauge theories can be derived from the duality of \cite{Aharony:1997gp}.  Starting from the electric theory $[\bm{0}, \bm{0}]a$ of section \ref{sec: aharony dual}, we can integrate out $n_f-s_2$ of the quarks $Q$ after giving them  a large negative real mass $m_0 < 0$. This gets us to a theory with $s_1=n_f$ antifundamentals, $s_2$ fundamentals, and CS level $k = - (s_1-s_2)/2$, which is precisely  the $[\bm{p}, \bm{0}]a$ case with $\bm{p}= s_1-s_2$.  This real mass deformation corresponds to a VEV for a background scalar from the flavor group, so we can easily identify the dual operation in the magnetic theory of \cite{Aharony:1997gp} and obtain the magnetic dual of our $[\bm{p}, \bm{0}]a$ theory. Integrating out chiral superfields with real masses also generates Chern-Simons level for the flavor group. For any two Abelian factors $U(1)_i$ and $U(1)_j$ which couple to the fermions $\psi$ with charges $q_i$, $q_j$, we generate a Chern-Simons level
\be\label{formula for shift of CS levels}
\delta  k_{ij} = \frac{1}{2} \sum_{\psi}\;  q_i(\psi)\,q_j(\psi)\; \text{sign}(M[\psi])\, .
\ee

Let us denote by $Q_{\beta}$ ($\beta=s_2+1, \cdots, s_1$) the $s_1-s_2$ massive quarks of real mass $M[Q_{\beta}]=m_0$, while the rest of the quarks are massless. From (\ref{real masses for Q Qt in general}), we must have
\be\label{limit for k2 theory}
\tilde{m}_a =0\, ,\qquad m_{b\neq \beta} = \frac{s_1-s_2}{s_1} m_0\, ,\qquad m_{\beta} = -\frac{s_2}{s_1} m_0\, , \qquad m_A=x_i= \frac{s_1-s_2}{2s_1}m_0\, ,
\ee
Remark that we also need to shift the real scalar for the diagonal $U(1)$ in the gauge group $U(n)$.
In this particular vacuum, the effective FI term is
\be
\xi_{eff}= \xi + \frac{s_1-s_2}{2} |m_0|\, ,
\ee
and this should vanish for the vacuum to be supersymmetric. Therefore we need to turn on a bare FI term $\xi= \frac{s_1-s_2}{2}m_0$. Denote by $F_{\beta}$ the $U(1)_{\beta}$ in the Cartan of $U(s_1-s_2) \subset U(s_1)$ under which $Q_{\beta}$ are charged, and by $F_b$ the Cartan of the remaining $U(s_2)$ group. Let us define the symmetry $U(1)_{\cF}$ such that
\be\label{def cF from mass}
M[\phi] = \cF[\phi] \, m_0\, ,
\ee
for any field $\phi$, when (\ref{limit for k2 theory}) holds. It is
\be\label{def of cF for [p,0] theory}
\cF\equiv \frac{s_1-s_2}{s_1} \sum_{b\neq\beta}  F_b - \frac{s_2}{s_1}\sum_{\beta} F_{\beta} +\frac{s_1-s_2}{2s_1} A_{[\bm{0},\bm{0}]} + \frac{s_1-s_2}{2}M + \frac{s_1-s_2}{2s_1}(Q_{e}+Q_{m}) \, ,
\ee
where $Q_{e}$ is the charge under the diagonal $U(1)$ in the $U(n)$ gauge groups, while $Q_{m}$ corresponds to the magnetic gauge group $U(m)$ (in the Aharony dual theory). In this formula $A_{[\bm{0},\bm{0}]}$ denotes the $U(1)_A$ symmetry of the $[\bm{0},\bm{0}]a$ theory we start with. Let us define a new $U(1)_A$ symmetry
\be
A_{[\bm{p},\bm{0}]}= A_{[\bm{0},\bm{0}]} +\sum F_{\beta}\, ,
\ee
such that $A[Q_b]=1$ for the light quarks ($b\neq \beta$), and $A[Q_{\beta}]=0$.
Equivalently, if we break the $SU(s_1)$ flavor group acting on the $Q$'s into
$SU(s_2)\times SU(s_1-s_2)\times U(1)_B$, where $B[Q_b]= s_1-s_2$ for $b \neq \beta$, and $B[Q_{\beta}]=-s_2$, we have
\be\label{def Ak2 from A,B,g}
A_{[\bm{p},\bm{0}]} = \frac{s_1+s_2}{2s_1}A_{[\bm{0},\bm{0}]} +\frac{1}{s_1}B -\frac{s_1-s_2}{2s_1}(Q_e+Q_m)\, .
\ee
This definition makes clear that the new axial symmetry is defined through a mixing with the gauge symmetry.
Turning on a real mass $m_{\tilde{A}}$ for the $A_{[\bm{p},\bm{0}]}$-symmetry corresponds to
\be\label{Atilde in term of A, B,g}
m_{A_{[\bm{0},\bm{0}]}}= \frac{s_1+s_2}{2s_1} m_{\tilde{A}}\, , \qquad m_B = \frac{1}{s_1}m_{\tilde{A}}\, , \qquad x_i = -\frac{s_1-s_2}{2s_1} m_{\tilde{A}}\, ,
\ee
including a finite shift of the scalar $\sigma$ in the vector multiplet.
We also define a new trial R-symmetry in the chiral theory
\be\label{redef R sym}
R_{[\bm{p},\bm{0}]} = R_{[\bm{0},\bm{0}]}- \frac{s_1-s_2}{2} M\, ,
\ee
assigning the same charge to the (bare) monopole operators $T_{[p,0]}$, $\tilde{T}_{[p,0]}$ of the IR theory.

In the electric theory $[\bm{0},\bm{0}]a$, we have
\be
\begin{array}{c|c|ccc|ccc|c}
    &  U(n)&  SU(s_1) & F_b  & F_{\beta} & A_{[\bm{p},\bm{0}]}& M & R_{[\bm{p},\bm{0}]} & \cF \\
\hline
\tilde{Q}^a     & \bm{\overline{n}} & \bm{s_1} & 0 & 0   &1  & 0   & 0   & 0 \\
Q_{b'}        & \bm{n} & \bm{1} & -\delta_{b'b}          & 0   &1  & 0   & 0   & 0 \\
\hline
Q_{\beta'}  & \bm{n} & \bm{1} & 0          & -\delta_{\beta'\beta}  &0   & 0   & 0   & 1 \\
\end{array}
\ee
The only microscopic fields charged under $\cF$ are $Q_{\beta}$, by definition. Integrating out the $s_1-s_2$ matter multiplets yields a CS level $k=-(s_1-s_2)/2$ for the gauge group and a global CS term for our choice of R-symmetry,
\be\label{KRR elec 01}
k_{RR}= -\frac12 n(s_1-s_2)\, .
\ee
There also are some more subtle global CS terms due to the definitions we gave for the axial and R-symmetries.
Let $A$ be the $U(n)$ gauge field, and $A^{(M)}$, $A^{(A)}$ and $A^{(R)}$ the background gauge fields for the $U(1)_M$, $U(1)_A$ and $U(1)_R$ symmetries, respectively. The FI parameter $\xi$ is the lowest component of the linear multiplet $\Sigma_M$ containing $dA^{(M)}$, so that
\be
\int d^3 x \int d^4\theta\, \Sigma_M \Tr(V) = \int d^3 x\, \xi \Tr(D) +  \int  dA^{(M)} \wedge \Tr(A)\, + \cdots
\ee
The shifts in (\ref{def Ak2 from A,B,g}) and (\ref{redef R sym}) modifies this last coupling according to
\be\label{FI parameter}
dA^{(M)} \wedge \Tr(A) \, \rightarrow \,  d\Big(A^{(M)}+ \frac{s_1-s_2}{2} A^{(R)}\Big)\wedge\Tr \Big(A -\frac{s_1-s_2}{2s_1}A^{(A)} \, \unit \Big)\, ,
\ee
and thus generates the mixed CS levels
\be\label{elec CS level for MA, RA}
k_{MA} = -n\frac{s_1-s_2}{2s_1}\, , \quad\qquad k_{RA} = -n \frac{(s_1-s_2)^2}{4s_1}\, ,
\ee
where the factor of $n$ comes from the trace over the gauge group.

In the magnetic theory of Aharony, we should integrate out all fields charged under $\cF$ (\ref{def of cF for [p,0] theory}). The charges of the various fundamental fields are summarized in table \ref{Table: charges magnetic theory for [p,0]a}.
Remark that the monopole operator $T$ survives as a singlet in the dual theory: half of the Coulomb branch of the electric theory survives. In the IR the magnetic theory is a $U(m)$ theory ($m= s_1-n$) at CS level $k_m= - k= (s_1-s_2)/2$. There are also various global Chern-Simons terms from integrating out, together with a contribution similar to (\ref{elec CS level for MA, RA}).
Since we are free to add the same global CS interactions to both sides of the duality, to state the final result we should express all global Chern-Simons levels as relative levels $\Delta k_{IJ} \equiv k^{magn}_{IJ} - k^{elec}_{IJ}$ between the magnetic and the electric description.
We also express everything in term of $m$, $s_1$, $s_2$ and $k$.
\begin{table}
\begin{center}
\be\nn
\begin{array}{c|c|ccc|ccc|c}
    &  U(m)&  SU(s_1) & F_b  & F_{\beta}  & A_{[\bm{p},\bm{0}]}& M & R_{[\bm{p},\bm{0}]} & \cF \\
\hline
\tilde{q}_a     & \bm{m} & \bm{\overline{s_1}} & 0 & 0   &-1  & 0   & 1   & 0 \\
q^{b'}       & \bm{\overline{m}} & \bm{1} & \delta_{b'b}          & 0   &-1  & 0   & 1   & 0 \\
M\ud{a}{b'}  & \bm{1} & \bm{s_1} & -\delta_{b'b}        & 0 &2    & 0   & 0   & 0 \\
T  & \bm{1} & \bm{1} & -\frac12 & -\frac12          & -\frac{s_1+s_2}{2}  & 1   & \frac{s_1+s_2}{2}-n+1     & 0 \\
\hline
q^{\beta'}  & \bm{\overline{m}} & \bm{1} & 0          & \delta_{\beta'\beta}  & 0   & 0   & 1   & -1 \\
M\ud{a}{\beta'}  & \bm{1} & \bm{s_1} & 0         & -\delta_{\beta'\beta}    & 1 & 0   & 0   & 1 \\
\tilde{T}  &  \bm{1}& \bm{1} & -\frac12 & -\frac12           & -\frac{s_1+s_2}{2}   & -1   &  \frac{3s_1-s_2}{2}-n+1   & -(s_1-s_2)
\end{array}
\ee
\end{center}
\caption{\small Charges for the fundamental fields in the magnetic theory of Aharony. The fields charged under $\cF$ (lower part of the table) are massive and integrated out in order to find the Seiberg dual of the $[\bm{p}, \bm{0}]a$ theory.}\label{Table: charges magnetic theory for [p,0]a}
\end{table}

The \textbf{electric} theory is a $U(n)_k$ gauge theory having matter content $(s_1, s_2)$ with $s_1 > s_2$, where the CS level is
\be
k= -\frac12 (s_1-s_2)\, .
\ee

The \textbf{magnetic} theory is a $U(s_1-n)_{-k}$ gauge theory with matter content $(s_2, s_1)$ plus $s_1s_2$ singlets $M^a_b$ and a singlet $T$, coupled by a superpotential
\be
W= \tilde{q}M q + t\tilde{T}\, ,
\ee
where $t$ is the gauge-invariant monopole operator of the magnetic gauge group. Moreover there are relative global CS levels
\be\label{[p,0]a global CS 1}
\Delta k_{SU(s_1)} = k\, , \qquad \qquad \Delta k_{SU(s_2)} = 0
\ee
for the non-Abelian part of the flavor group $SU(s_1)\times SU(s_2)$ and
\bea\label{[p,0]a global CS 2}
& \Delta k_{AA}=ks_1 +\frac18(s_1+s_2)^2\, , &\quad\qquad & \Delta k_{AM}= -k +\frac14 (s_1+ s_2) \\
& \Delta k_{MM}= \frac12 \, , &\quad\qquad & \Delta k_{AR}= -k(s_1+k)-\frac14(s_1+s_2)(m-k) \\
& \Delta k_{RR}= km+\frac12(m-k)^2 \, , &\qquad\quad & \Delta k_{MR}= -\frac12 (m-k)\, .
\eea
for the Abelian flavor symmetries $U(1)_A\times U(1)_M\times U(1)_R$, where $m= s_1-n$.

The above formulas extend to the case $s_2=0$, ignoring the CS levels involving $U(1)_A$.

\subsection{Dual of the $[\bm{p}, \bm{q}]a$ theory}
\label{subsec: FT analysis of [k,l] duals}

To go to a $[\bm{p}, \bm{q}]a$  theory with $pq>0$, we start again from the non-chiral theory $[\bm{0},\bm{0}]a$, and we integrate out $n_f-s_1$ of the antiquarks $\tilde{Q}$ (denoted $\tilde{Q}^\alpha$, $\alpha= s_1+1, \cdots, n_f$) together with $n_f-s_2$ of the quarks $Q$ (denoted $Q_{\beta}$, $\beta= s_2+1, \cdots, n_f$), with $M[\tilde{Q}^\alpha]=M[Q_{\beta}]= m_0 < 0$. For this we take the mass parameters and FI terms
\bea\label{limit to get kl theory}
& \tilde{m}_{a} = -\frac{n_f-s_1}{n_f} m_0 \, ,&& \tilde{m}_{\alpha} = \frac{s_1}{n_f} m_0\, ,&\quad & m_b = \frac{n_f-s_2}{n_f}m_0\, ,& \quad &m_{\beta} = -\frac{s_2}{n_f} m_0& \\
& m_A= \frac{2n_f-s_1-s_2}{2n_f}m_0\, ,&  & x_i= \frac{s_1-s_2}{2n_f} m_0\, , &  &\xi= \frac{1}{2}(s_1-s_2) m_0\, , &
\eea
with $m_0\rightarrow -\infty$. Similarly to the previous subsection, we define a symmetry $\cF$ such that (\ref{def cF from mass}) holds when the masses (\ref{limit to get kl theory}) are turned on,
and the new axial and R-symmetries
\be\label{Akl def 0}
A_{[\bm{p},\bm{q}]}= A_{[\bm{0},\bm{0}]} -\sum_{\alpha} \tilde{F}_{\alpha} +\sum_{\beta} F_{\beta}\, ,\qquad \quad\qquad
R_{[\bm{p},\bm{q}]} = R_{[\bm{0},\bm{0}]}- \frac{s_1-s_2}{2} M\, .
\ee
In the electric $[\bm{0},\bm{0}]a$ theory, we have
\be
\begin{array}{c|c|cccc|ccc|c}
    &  U(n)&   \tilde{F}_a  & \tilde{F}_{\alpha} & F_b & F_{\beta} & A_{[\bm{p},\bm{q}]}& M & R_{[\bm{p},\bm{q}]} & \cF \\
\hline
\tilde{Q}^a               & \bm{\overline{n}} & 1 & 0 & 0  &0 &1  & 0   & 0  & 0 \\
Q_{b}                     & \bm{n}            & 0& 0          & -1 &0 &1   & 0   & 0   & 0 \\
\hline
\tilde{Q}^\alpha       & \bm{\overline{n}}  & 0 & 1  & 0  &0 &0  & 0   & 0   & 1 \\
Q_{\beta}                     & \bm{n}            & 0& 0          & 0 &-1 &0   & 0   & 0   & 1 \\
\end{array}
\ee
Integrating out $\tilde{Q}^\alpha$ and $Q_{\beta}$ yields a CS level $k=-n_f + \frac{s_1+s_2}{2}$ for the $U(n)$ gauge group, together with global CS terms. This leads us to the $[n_f-s_2, n_f-s_1]a$ field theory.
The charges of the fields in the magnetic theory are reproduced in table \ref{Table: charges magnetic theory for [p,q]a}.

\begin{table}
\begin{center}
\be\nn
\begin{array}{c|c|cccc|ccc|c}
    &  U(m)& \tilde{F}_a  & \tilde{F}_{\alpha} & F_b & F_{\beta} & A_{[\bm{p},\bm{q}]}& M & R_{[\bm{p},\bm{q}]} & \cF \\
\hline
\tilde{q}_{a}     & \bm{m} & -1& 0 & 0  &0 &-1  & 0   & 1   & 0 \\
q^b  & \bm{\overline{m}} & 0 & 0          & 1 &0 & -1   & 0   & 1   & 0 \\
M\ud{a}{b}  & \bm{1} & 1  & 0  & -1 &0 &2   & 0   & 0   & 0 \\
\hline
\tilde{q}_{\alpha}      & \bm{m} & 0 & -1    &  0 &0 &0  & 0   & 1   & -1 \\
q^{\beta}  & \bm{\overline{m}} & 0 & 0          & 0 &-1 & 0   & 0   & 1   & -1 \\
M\ud{\alpha}{b}  & \bm{1} & 0 & 1   &  -1  & 0 & 1 & 0   & 0   & 1 \\
M\ud{a}{\beta}  & \bm{1} & 1 & 0   &  0  & -1 & 1 & 0   & 0   & 1 \\
M\ud{\alpha}{\beta}  & \bm{1} & 0 & 1   &  0  & -1 & 0 & 0   & 0  & 2 \\
T  &  \bm{1}&-\frac12 & -\frac12 & -\frac12     &  -\frac12 & -\frac12(s_1+s_2)  & 1   & n_f-n-\frac{s_1-s_2}{2}+1   & -(n_f-s_1)\\
\tilde{T}  &  \bm{1}&-\frac12 & -\frac12 & -\frac12     &  -\frac12 & -\frac12(s_1+s_2)  & -1   & n_f-n+\frac{s_1-s_2}{2}+1   & -(n_f-s_2)
\end{array}
\ee
\end{center}
\caption{\small Charges for the fundamental fields in the magnetic theory. The fields charged under $\cF$ are integrated out to obtain the Seiberg dual of the $[\bm{p}, \bm{q}]a$ theory.}\label{Table: charges magnetic theory for [p,q]a}
\end{table}

The \textbf{electric} SQCD-like theory is a $U(n)_k$ CS theory with matter content $(s_1, s_2)$, with
\be
k < -\frac12|s_1-s_2| \;.
\ee

The \textbf{magnetic} theory is a $U(m)_{-k}$ CS theory, where
\be
m = \frac12(s_1+s_2) -k -n \;,
\ee
with matter content $(s_2, s_1)$ along with $s_1s_2$ singlets $M$, and superpotential $W=\tilde{q}Mq$. In addition there are relative global CS interactions at level
\be\label{CS level for nAG in pq case}
\Delta k_{SU(s_1)} = \frac12 \Big( k-\frac{s_1-s_2}{2} \Big)\, , \qquad \quad \Delta k_{SU(s_2)} = \frac12 \Big( k+\frac{s_1-s_2}{2} \Big)
\ee
for the non-Abelian flavor symmetry group and
\bea
& \Delta k_{AA}= \frac12 k(s_1+s_2) +s_1s_2 \, , &\qquad\quad & \Delta k_{AM}= \frac12(s_1-s_2) \\
& \Delta k_{MM}= 1 \, , &\qquad\quad                                           & \Delta k_{AR}= -\frac12(k+m)(s_1+s_2) \\
& \Delta k_{RR}= \frac12 \big( (k+m)^2 +m^2 \big) +\frac18 (s_1-s_2)^2  \, , &\qquad\quad & \Delta k_{MR}=-\frac12(s_1-s_2)
\eea
for the Abelian symmetries.

\subsection{Dual of the $[\bm{p}, \bm{q}]^*a$ theory}
\label{subsec: FT analysis of [k,l]* duals}

To obtain a $[\bm{p}, \bm{q}]^*a$ theory, we start again from a $[\bm{0}, \bm{0}]a$ theory with $n_f$ flavors. We now keep the $n_f=s_1$ antiquarks $\tilde{Q}$ massless, while integrating out $s_1-\ts$ of the quarks (denoted $Q_{\beta}$) with a negative mass and $\ts-s_2$ of the quarks (denoted $Q_{\gamma}$) with a positive mass. For this we take the limit $m_0\rightarrow -\infty$, scaling
\bea\label{limit to kl* from 22}
& \tilde{m}_a =0\, ,&\quad & m_{b} = \frac{s_1+s_2-2\ts}{s_1} m_0\, , \\
 & m_{\beta} = -\frac{2\ts-s_2}{s_1} m_0\, ,&\quad & m_{\gamma} =\frac{2s_1+s_2-2\ts}{s_1}m_0  \, , \\
 & m_A=x_i= \frac{s_1+s_2-2\ts}{2s_1}m_0 & \, , \quad & \xi = \frac12 (s_1-s_2) m_0\, .
\eea
We define a symmetry $\cF$ as before,
and the new axial and R-symmetries
\be
A_{[\bm{p},\bm{q}]^*}= A_{[\bm{0},\bm{0}]} +\sum_{\beta}^{s_1-\ts} F_{\beta}+\sum_{\gamma}^{\ts-s_2} F_{\gamma}\, , \quad \qquad \quad
R_{[\bm{p},\bm{q}]^*} = R_{[\bm{0},\bm{0}]}- \frac{s_1+s_2-2\ts}{2} M\, .
\ee
In the electric theory we have
\be
\begin{array}{c|c|cccc|ccc|c}
    &  U(n)&  SU(s_1) & F_b  & F_{\beta} & F_{\gamma} & A_{[\bm{p},\bm{q}]^*}& M & R_{[\bm{p},\bm{q}]^*} & \cF \\
\hline
\tilde{Q}^a     & \bm{\overline{n}} & \bm{s_1} & 0 & 0  &0 &1  & 0   & 0   & 0 \\
Q_{b}        & \bm{n} & \bm{1} & -1     & 0  &0 &1  & 0   & 0   & 0 \\
\hline
Q_{\beta}  & \bm{n} & \bm{1} & 0          & -1 &0 &0   & 0   & 0   & 1 \\
Q_{\gamma}  & \bm{n} & \bm{1} & 0          & 0 &-1 &0   & 0   & 0   & -1
\end{array}
\ee
Integrating out the heavy quarks generates a CS level $k=\ts -(s_1+s_2)/2$ for the $U(n)$ gauge group and leads us to the $[s_1-\ts,\ts-s_2]^*a$ theory.
The charges for the magnetic theory are given in table \ref{Table: charges magnetic theory for [p,q]*a}.

\begin{table}
\begin{center}
\be\nn
\begin{array}{c|c|cccc|ccc|c}
    &  U(m)&  SU(s_1) & F_b  & F_{\beta} & F_{\gamma} & A_{[\bm{p},\bm{q}]^*}& M & R_{[\bm{p},\bm{q}]^*} & \cF \\
\hline
\tilde{q}_a     & \bm{m} & \bm{\overline{s_1}} & 0 & 0  &0 &-1  & 0   & 1   & 0 \\
q^{b}       & \bm{\overline{m}} & \bm{1} & 1          & 0  &0 &-1  & 0   & 1   & 0 \\
M\ud{a}{b}  & \bm{1} & \bm{s_1} & -1        & 0 &0 &2   & 0   & 0   & 0 \\
\hline
q^{\beta}  & \bm{\overline{m}} & \bm{1} & 0          & 1 &0 & 0   & 0   & 1   & 1 \\
q^{\gamma}  & \bm{\overline{m}} & \bm{1} & 0          & 0 &1 & 0   & 0   & 1   & -1 \\
M\ud{a}{\beta}  & \bm{1} & \bm{s_1} & 0         & -1  & 0 & 1 & 0   & 0   & 1 \\
M\ud{a}{\gamma}  & \bm{1} & \bm{s_1} & 0         & 0  & -1 & 1 & 0   & 0  & -1 \\
T  & \bm{1} & \bm{1} & -\frac12 & -\frac12         & -\frac12  & -\frac{s_1+s_2}{2}  & 1  &  \frac{s_1+2\ts -s_2-2n}{2}+1   & \ts-s_2\\
\tilde{T}  & \bm{1} & \bm{1} & -\frac12 & -\frac12& -\frac12  & -\frac{s_1+s_2}{2}  & -1   &  \frac{3s_1-2\ts +s_2-2n}{2}+1  & -(s_1-\ts)
\end{array}
\ee
\end{center}
\caption{\small Charges for the fundamental fields in the magnetic theory. The fields charged under $\cF$ are integrated out to obtain the Seiberg dual of the $[\bm{p}, \bm{q}]^*a$ theory.}\label{Table: charges magnetic theory for [p,q]*a}
\end{table}

The \textbf{electric} theory is a $U(n)_k$ CS theory with matter content $(s_1,s_2)$ with $s_1>s_2$ and
\be
|k|<\frac12|s_1-s_2| \;.
\ee

The \textbf{magnetic} theory is a $U(s_1-n)_{-k}$ theory with matter content $(s_2, s_1)$ together with $s_1s_2$ singlets $M$, and superpotential $W=\tilde{q}Mq$. There are relative global CS terms
\be
\label{CS level for nAG in pq* case}
\Delta k_{SU(s_1)} = k \;,\qquad \quad \Delta k_{SU(s_2)} = 0
\ee
for the non-Abelian flavor symmetry group and
\bea
& \Delta k_{AA}= ks_1 \, , &\qquad\quad & \Delta k_{AM}= s_1 \\
& \Delta k_{MM}= 0 \, , &\qquad\quad    & \Delta k_{AR}= 0 \\
& \Delta k_{RR}= -km  \, , &\qquad\quad & \Delta k_{MR}= -m
\eea
for the Abelian symmetries, with $m=s_1-n$.

\subsection{Summary of the $U(n)_k$ dualities}
The missing cases can be obtained from the ones above by parity (P) and/or charge conjugation (C).
Firstly, if we go from a $U(n)_k$ theory with matter content $(s_1,s_2)$ and with $k < 0$ to another such theory with $k>0$, the global CS levels change according to
\be\label{def P tranform}
\text{P}\, :\, \quad k_{\text{global}}= f(k, s_1, s_2, m) \, \quad \rightarrow\quad \, k_{\text{global}}= - f(-k, s_1, s_2, m)\, .
\ee
This is our P transformation. Secondly, if we go from a theory with $s_1 > s_2$ to a theory with $s_2 > s_1$, the net effect on the global CS levels is obtained by exchanging $s_1$ and $s_2$:%
\footnote{We loosely call this operation and the previous one $C$ and $P$, but they are really the $C$ and $P$ transformations together with a relabeling of the parameters.}
\be\label{def C transform}
\text{C}\, :\, \quad k_{\text{global}}= f(k, s_1, s_2, m) \, \quad\rightarrow\quad \, k_{\text{global}}= f(k, s_2, s_1, m)\, .
\ee

The \textbf{electric} theories $[\bm{p},\bm{q}]a$ and $[\bm{p},\bm{q}]b$ -- that we call \emph{minimally chiral} -- have gauge group $U(n)_k$ and matter content $(s_1,s_2)$ with
\be
|k| > \frac12 |s_1 - s_2| \geq 0 \;.
\ee
The dual \textbf{magnetic} theories have gauge group $U(m)_{-k}$, where
\be
m = \frac12(s_1 + s_2) +|k| - n \;,
\ee
with matter content $(s_2,s_1)$ along with $s_1s_2$ singlets $M$ and superpotential $W = \tilde q M q$. The relative global CS levels are
\bea\label{global CS level [p,q] case in general}
&\Delta k_{SU(s_1)} = \frac12 \Big( k + \sign(k)\, \frac{s_1 - s_2}2 \Big)\, ,  &\qquad
&\Delta k_{SU(s_2)} = \frac12 \Big( k - \sign(k)\, \frac{s_1 - s_2}2 \Big)\, , \\
&\Delta k_{AA} = \frac12 k(s_1+s_2) - \sign(k)\, s_1s_2\, , &\qquad
&\Delta k_{AM} = - \sign(k) \frac{s_1 - s_2}2\, , \\
&\Delta k_{MM} = -\sign(k)\, , &\qquad
&\Delta k_{AR} =  \sign(k)\frac{(m - |k|)(s_1 + s_2)}2\, , \\
\quad &\qquad
&&\Delta k_{MR} =  \sign(k)\frac{s_1 - s_2}2\, ,
\eea
\bea\nn
& \Delta k_{RR} = -\sign(k) \Big( \frac{(m - |k|)^2 + m^2}2 + \frac{(s_1 - s_2)^2}8 \Big)\, . &\;\qquad\qquad\qquad\quad\qquad&
\eea

The \textbf{electric} theories $[\bm{p},\bm{q}]^*a$ and $[\bm{p},\bm{q}]^*b$ -- that we call \emph{maximally chiral} -- have gauge group $U(n)_k$ and matter content $(s_1,s_2)$ with
\be
0 \leq |k| < \frac12 |s_1 - s_2| \;.
\ee
The dual \textbf{magnetic} theories have gauge group $U(m)_{-k}$, where
\be
m = \max(s_1,s_2) - n \;,
\ee
with matter content $(s_2,s_1)$ along with $s_1s_2$ singlets $M$ and superpotential $W = \tilde q M q$. The relative global CS levels are
\bea
&\Delta k_{SU\big( \max(s_1,s_2) \big)} = k\, , \qquad\qquad\qquad &
&\Delta k_{SU\big( \min(s_1,s_2) \big)} = 0\, , \\
&\Delta k_{AA} = k \max(s_1,s_2)\, , &
&\Delta k_{AM} = \max(s_1,s_2)\, , \\
&\Delta k_{MM} = 0 \, ,&
&\Delta k_{AR} = 0 \, ,\\
&\Delta k_{RR} = -k\,m\, , &
&\Delta k_{MR} = -m \;.
\eea

Similarly, one can work out the limiting cases $[\bm{0},\bm{p}]b$, $[\bm{0}, \bm{p}]a$ and $[\bm{p},\bm{0}]b$ from the $[\bm{p},\bm{0}]a$ case of section \ref{subsec: dual of [p,0]a theory}. In the $[\bm{0},\bm{p}]b$ case, the electric theory is $U(n)_k$ with matter content $(s_1, s_2)$ and $k = \frac12(s_1-s_2)>0$, and the relative global CS levels are obtained from (\ref{[p,0]a global CS 1})-(\ref{[p,0]a global CS 2}) by a P transformation (\ref{def P tranform}). In the $[\bm{0},\bm{p}]a$ case, the electric theory $U(n)_k$ has $k = \frac12(s_1-s_2)<0$, and the global CS levels are obtained from (\ref{[p,0]a global CS 1})-(\ref{[p,0]a global CS 2}) by a C transformation (\ref{def C transform}). Finally, the $[\bm{p},\bm{0}]b$ case corresponds to an electric theory with $k= -\frac12 (s_1-s_2)>0$, and it is obtained from the $[\bm{p},\bm{0}]a$ case by a CP transformation.


\section{Consequences for Chern-Simons quivers}
\label{sec: quivers}

In recent years it was realized that M2-branes at Calabi-Yau singularities can be described by $\cN=2$ Chern-Simons  quivers \cite{Aharony:2008ug}. Generally such a quiver has gauge group $U(n_1)\times \cdots\times U(n_G)$ with some CS interactions, and fields $X_{ij}$ in bifundamental representations coupled together by a superpotential \cite{Imamura:2008qs, Hanany:2008cd, Martelli:2008si, Franco:2009sp, Aganagic:2009zk, Davey:2009sr, Tomasiello:2010zz}.  The M2-brane quivers can also include so-called flavors (fields in the fundamental/antifundamental representation of some of the $U(n_i)$ factors) \cite{Gaiotto:2009tk, Benini:2009qs, Jafferis:2009th}.  While most of the literature focussed on the case with equal ranks $n_i=N$, it was argued recently that this is not the case for rather generic M-theory backgrounds \cite{Benini:2011cm}.

Seiberg-like dualities for M2-brane quivers were studied in \cite{Aharony:2008gk, Amariti:2009rb, Jensen:2009xh}, but a thorough understanding of Seiberg-like dualities for chiral quivers is still missing.
While we hope to report some progress in that direction in the near future, in this section we briefly discuss some obvious consequences of the dualities of section \ref{sec: SD for chiral SQCD--general} for chiral quivers.

\begin{figure}[t]
\begin{center}
\subfigure[\small Local neighborhood in a CS quiver.]{
\includegraphics[height=4cm]{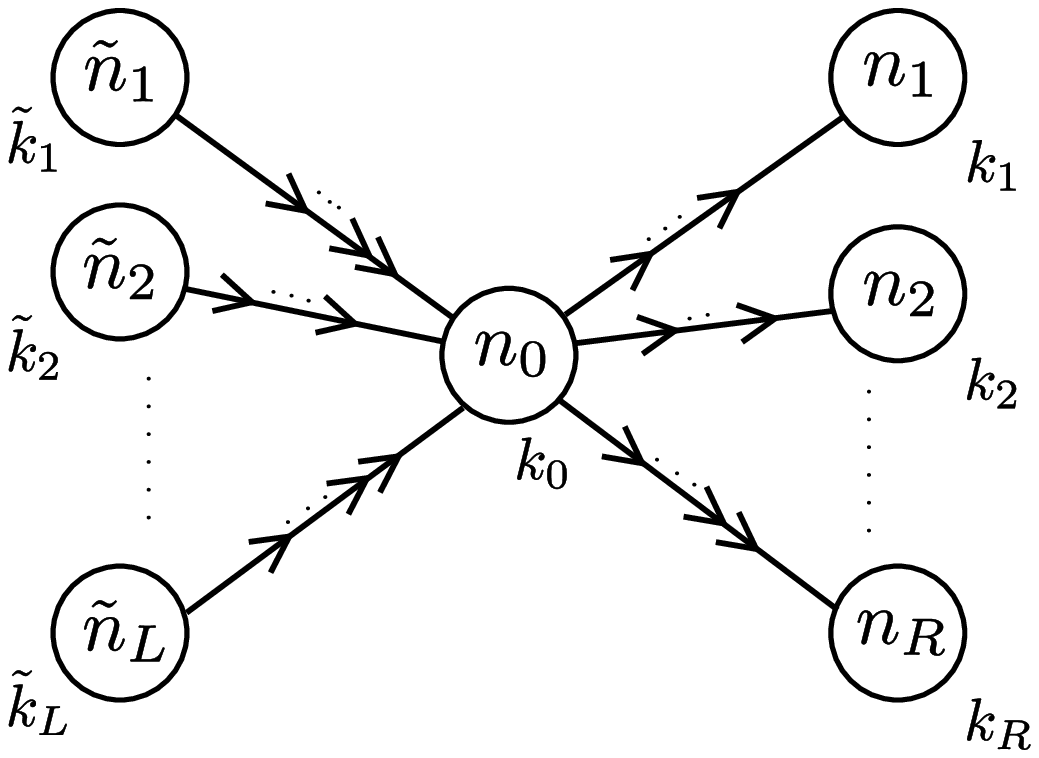}
\label{fig:quiver elec}}
\qquad \qquad
\subfigure[\small The Seiberg dual neighborhood.]{
\includegraphics[height=4cm]{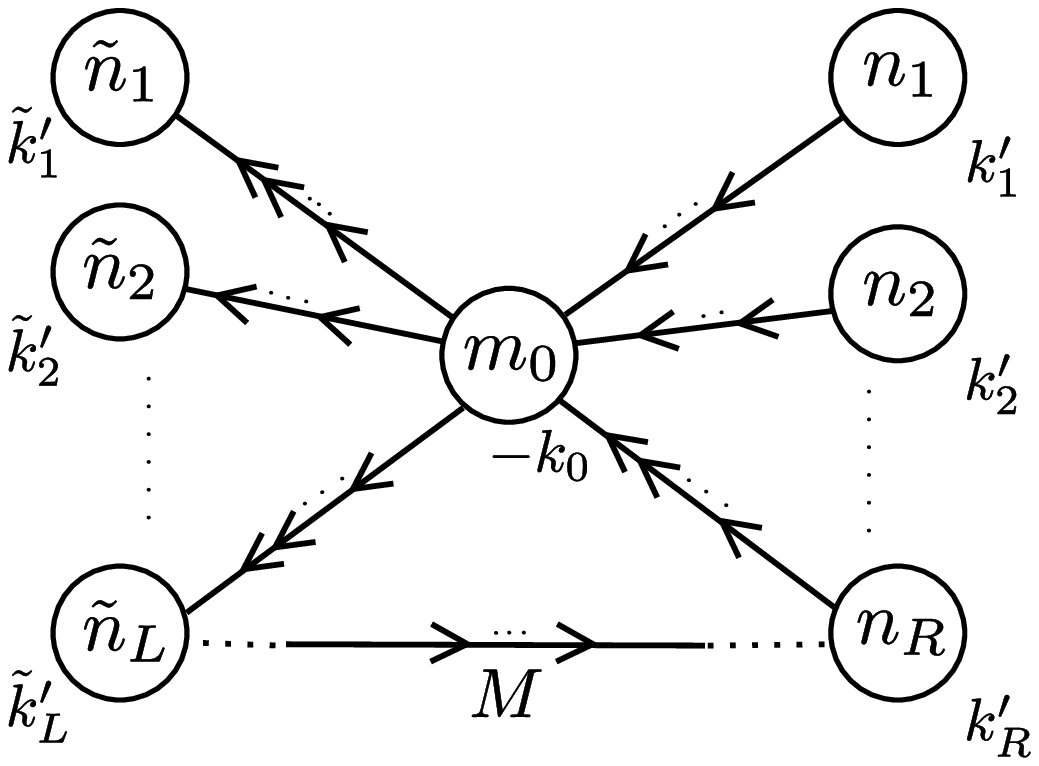}
\label{fig:quiver magn}}
\end{center}
\caption{\small Seiberg-like duality acting on a quiver. Remark that the dual quiver also contains some mesonic fields $M$, although generally many of the mesons will be massive due to the superpotential.}\label{Fig: SD on quiver}
\end{figure}

Let us consider a generic quiver of the kind described above. We pick a node with gauge group $U(n_0)$ and Chern-Simons level $k_0$, and we consider its direct neighborhood, as shown in figure \ref{fig:quiver elec}. The nodes connected to the node $0$ with incoming arrows have gauge groups $U(\tilde{n}_i)$, $i=1, \cdots, L$, and the nodes connected to node $0$ with outgoing arrows have gauge groups $U(n_j)$, $j=1, \cdots, R$. Some of these contiguous nodes could be global symmetries, corresponding to flavors. The contiguous nodes have CS levels $\tilde{k}_i$ and $k_j$, respectively. Let us denote by $A_{i0}$ the number of bifundamental fields between $U(\tilde{n_i})$ and $U(n_0)$, and by $A_{0j}$ the number of bifundamental fields between $U(n_0)$ and $U(n_j)$. The total numbers of antifundamental and fundamentals fields for $U(n_0)$ is
\be
s_1\,\equiv \,  \sum_{i=1}^F \tilde{n}_i \, A_{i0}  \, , \qquad\qquad \qquad  s_2\,\equiv \,  \sum_{j=1}^R  A_{0j}\, n_j\, .
\ee
The big flavor group $SU(s_1)\times SU(s_2)$ of chiral SQCD splits according to%
\footnote{In the following we ignore various $U(1)$ factors. In general the $U(1)$ and $SU(n)$ part of the $U(n)$ gauge groups can have different CS levels, and one should keep track of how the CS terms for the Abelian sector transform under duality. A full discussion of this is beyond the goal of the present section.}
\be
SU(s_1) \rightarrow \prod_i SU(\tilde{n}_i) \times SU(A_{i0})\, , \qquad SU(s_2) \rightarrow \prod_j SU(n_j) \times SU(A_{0j})\, ,
\ee
where the factors $SU(\tilde{n}_i)$, $SU(n_j)$ are gauged (unless the node $i$ or $j$ stands for a flavor group), while the factors $SU(A_{i0})$, $SU(A_{0j})$ are global symmetries, which can be further reduced due to the superpotential.

Performing a Seiberg-like duality on node $0$ leads to the situation of figure \ref{fig:quiver magn}. All arrows connected to node $0$ are reversed, and various mesonic fields are generated. Some of the mesons may have a complex mass, in which case they should be integrated out. While the rank $n_0$ is the only one which changes under Seiberg-like duality on node $0$, the CS levels of the contiguous nodes are in general affected. We have two distinct situations, corresponding to cases $[\bm{p}, \bm{q}]$ and $[\bm{p}, \bm{q}]^*$ of section \ref{sec: SD for chiral SQCD--general}.

\subsection{Minimally chiral case}

The first case is a direct generalization of the non-chiral case studied in \cite{Amariti:2009rb, Jensen:2009xh}. If $|k_0|\geq  \frac12 |s_1-s_2|$, the Seiberg dual has a gauge group $U(m_0)$ at level $-k_0$ for node $0$, with
\be
m_0 = \frac{s_1+s_2}{2}+ |k_0|- n_0\, .
\ee
The CS levels of the contiguous nodes are then
\be\nn
\tilde{k}_i'\, =\, \tilde{k}_i +\frac12 A_{i0}\left(k_0+ \sign(k_0)\frac{s_1-s_2}{2}\right)\, , \quad
k_j' \,=\, k_j +\frac12 A_{0 j}\left(k_0-  \sign(k_0) \frac{s_1-s_2}{2}\right)\, ,
\ee
which can be deduced from (\ref{CS level for nAG in pq case}) or (\ref{global CS level [p,q] case in general}).

\subsection{Maximally chiral case}

The ``maximally chiral'' case is when $|k_0| \leq \frac12|s_1-s_2|$. When $s_1 >s_2$, we have $U(m_0)$ at level $-k_0$ for node $0$, with
\be\label{max chiral SD quiv 01}
m_0 = s_1- n_0\, ,
\ee
while the CS levels of the contiguous groups are
\be\label{max chiral SD quiv 02}
\tilde{k}_i'\, =\, \tilde{k}_i + A_{i0}\, k_0\, , \qquad \qquad
k_j' \,=\, k_j \, .
\ee
This follows from (\ref{CS level for nAG in pq* case}). Similarly, if $s_2>s_1$, we have
\be\label{max chiral SD quiv 03}
m_0 = s_2- n_0\, ; \qquad \qquad
\tilde{k}_i'\, =\, \tilde{k}_i\, , \qquad \quad
k_j' \,=\, k_j  + A_{0j}\, k_0\, .
\ee
In four dimensional quivers, Seiberg duality of an interesting class of chiral quivers (del Pezzo quivers in particular) can be understood in term of so-called ``mutations'', which are mathematical operations which change the basis of fractional branes at a Calabi-Yau singularity \cite{Cachazo:2001sg, Feng:2002kk, Herzog:2004qw}. It turns out that the mutation prescription of \cite{Herzog:2004qw} exactly corresponds to the Seiberg-like duality for the maximally chiral case (\ref{max chiral SD quiv 01})-(\ref{max chiral SD quiv 02}) or (\ref{max chiral SD quiv 03}), for left or right mutations, respectively \cite{unpublished:cyril}.%
\footnote{In 4d we have $s_1=s_2$ to cancel the gauge anomalies, and there are no CS interactions, therefore the distinction we made between minimally and maximally chiral theories does not arise.}


\section{Dualities for $U(n)$ theories and $S^3_b$ partition function}
\label{sec: Un duality and S3 PF}

Recently it was understood that the partition function $Z_{S^3}$ for any $\cN=2$ superconformal theory on a 3-sphere can be reduced to an ordinary integral using localization techniques \cite{Kapustin:2009kz, Jafferis:2010un, Hama:2010av}. The $S^3$ partition function can be computed for any trial R-charge. We can further squash the $S^3$ into a $U(1)\times U(1)$ isometric hyperellipsoid $S^3_b$ with squashing parameter $b$ as in \cite{Hama:2011ea}  ($b=1$ corresponding to the round sphere).

Jafferis argued that $|Z_{S^3}|$ is extremized%
\footnote{In all known examples it is minimized.}
by the superconformal R-charge \cite{Jafferis:2010un}. We will apply Z-extremization to our theories in section \ref{sec: Z-min}. In this section we will derive identities between partition functions which provide strong checks of the proposed dualities.

\subsection{$S^3_b$ partition function for chiral $U(n)$ SQCD}
\label{sec: partitions functions and dualities}

Consider the partition function of the electric theory as defined at the beginning of section \ref{sec: SD for chiral SQCD--general}, namely $U(n)_k$ gauge theory with $(s_1,s_2)$ matter.
The localized partition function is an integral on the VEVs of the real scalar $\sigma= (x_1, \cdots, x_n)$ in the vector multiplet, all the others fields being frozen to zero expectation value. As a function of the real scalars in background vector multiplets (\ref{def real masses}), the partition function of the electric theory reads%
\footnote{Remark that our conventions agree with the ones of \cite{Jafferis:2010un} except for a sign flip of all CS levels. In the present conventions integrating out a charged fermion of real mass $m$ and charge $+1$ results in a CS level $\delta k=+\frac12 \sign(m)$ for the corresponding background gauge field.}
\begin{equation} \label{Z electric}
\begin{split}
Z_{U(n)_k}^{(s_1,\,s_2)} (\tilde{m}_a; m_b; & m_A; \xi) = \frac{1}{n !}  \int   \prod_{1 \leq i< l\leq n} \frac{1}{\Gamma_h(\pm (x_i-x_l))} \; \prod_{j=1}^{n} dx_j \; e^{-\pi i \,k\, x_j^2 -2 \pi i\, \xi\, x_j} \\
&\times \prod_{a=1}^{s_1} \Gamma_h(-x_j+\tilde{m}_a +m_A +\omega\, r_Q) \prod_{b=1}^{s_2} \Gamma_h(x_j-m_b +m_A+\omega\, r_Q) \;,
\end{split}
\end{equation}
where $\omega = i\frac{b+b^{-1}}{2}$ ($\omega=i$ for the round $S^3$).
The building block of the integrand is the hyperbolic Gamma function $\Gamma_h(z)\equiv \Gh(z;\omega_1,\omega_2)$, with periods $\omega_1=i b$ and $\omega_2=i/b$ understood from now on. Several useful formul\ae{} about $\Gamma_h$ and our notations are gathered in appendix \ref{app: Gammah}. In particular, we define $\Gamma_h(a\pm x)= \Gamma_h(a+x)\Gamma_h(a-x)$.
We are free to choose an R-symmetry $R$ so that $R[\tilde{Q}] = R[Q] \equiv r_Q$ by a mixing with the central $U(1)$ gauge symmetry.
The $U(1)_M$ symmetry can also mix with the R-charge (it corresponds to an imaginary part for the FI parameter $\xi$), but that only affects the R-charge of monopole operators. More generally, mixing a global symmetry $U(1)_I$ with the R-symmetry corresponds to giving an imaginary part to the real mass $m_I$ according to \cite{Jafferis:2010un, Hama:2010av, Hama:2011ea}
\be\label{mixing with R and omega}
m_I \rightarrow m_I + \omega\, a_I\, \qquad \Leftrightarrow  \qquad  R \rightarrow R + a_I Q_I\,,
\ee
where $Q_I$ is the  $U(1)_I$ charge.

Integrals such as (\ref{Z electric}) were studied in recent years by mathematicians, in particular by van de Bult \cite{vdbult_thesis}.
For the integers $n, s_1, s_2 \in \bZ_{\geq 0}$ and $t \in \bZ$ (unless $s_1=s_2=t=0$), van de Bult defines (Definition 5.3.17 in \cite{vdbult_thesis})
\begin{equation}\label{def Jnsst}
\begin{split}
J_{n,\, (s_1, \,s_2),\, t} (\mu; \nu; \lambda) \equiv &
\frac{1}{\sqrt{-\omega_1\omega_2}^n \, n!}  \int_{C^n}
\prod_{1 \leq i< l\leq n} \frac{1}{\Gamma_h \big( \pm(x_i-x_l) \big)}\,\\
&\times\, \prod_{j=1}^n \;dx_j\;  e^{\frac{\pi i}{2\omega_1\omega_2}(t x_j^2 + 2 \lambda x_j)} \, \prod_{a=1}^{s_1} \Gamma_h(\mu_a - x_j) \prod_{b=1}^{s_2} \Gamma_h(\nu_b + x_j) \;.
\end{split}
\end{equation}
The contour $C$ is a path that goes from $\re{x} = -\infty$ to $\re{x}= \infty$, goes below all the poles from $\Gamma_h(\mu-x)$ and above all the poles from $\Gamma_h(\nu+x)$; it can go to infinity on the left and right with some tilt so that (\ref{def Jnsst}) converges -- see \cite{vdbult_thesis} for the precise definition. In this paper we will always take $C$ to be the real axis.
We see that, for $\omega_1=i b$, $\omega_2=i b^{-1}$,
\be
 Z_{U(n)_k}^{(s_1,\,s_2)}(\tilde{m}_a; m_b; m_A; \xi) = J_{n,\, (s_1,\, s_2),\, 2k} (\mu; \nu; \lambda) \, ,
\ee
with
\be
\lambda= 2\xi\, , \qquad \mu_a= \tilde{m}_a+m_A+ \omega\, r_Q \, , \qquad \nu_b= -m_b+m_A+ \omega\, r_Q\, .
\ee
We have the constraints
\be\label{condition on nu mu}
\sum_{a=1}^{s_1} \mu_a=s_1 m_A\, , \qquad\quad   \sum_{b=1}^{s_2} \nu_b=  s_2 m_A\, .
\ee
Following closely the notation of \cite{vdbult_thesis}, we rename $Z_{U(n)_k}^{ (s_1,\, s_2)}$ according to the four cases (\ref{The chiral electric theories}):
\be\label{def Zelec in 4 cases}
Z_{U(n)_k}^{(s_1,\, s_2)} \, =\, \begin{cases} I_{n\, [p,q]a}^m & \qquad  \text{if} \qquad  k \leq - \frac{1}{2} |s_1-s_2| \\
                                   I_{n\, [p,q]b}^m & \qquad  \text{if} \qquad  k\geq +\frac{1}{2} |s_1-s_2| \\
                                  I_{n\, [p,q]^*a}^m & \qquad  \text{if} \qquad  |k|\leq \frac{1}{2} ( s_1-s_2) \\
                                     I_{n\, [p,q]^*b}^m & \qquad  \text{if} \qquad  |k|\leq  \frac{1}{2} (s_2-s_1)
\end{cases}
\ee
where the rank $m$ of the dual gauge group, is given in (\ref{m for various cases}), and $p$ and $q$ are related to the effective CS levels $k_\pm$ \eqref{k+-} as in \eqref{The chiral electric theories}.

\subsection{Aharony duality and real mass deformations}

Consider Aharony duality \cite{Aharony:1997gp} reviewed in section \ref{sec: aharony dual}. At the level of the partition function, it can be written (recall $n+m=n_f$)
\bea
\label{aharony duality ZeZ}
I_{n\, [0,0]a}^m (\mu; \nu; \lambda) = I_{m\, [0,0]b}^n (\omega-\nu; \omega-\mu; -\lambda)\, \Gamma_h \Big((m+1)\omega -n_f m_A \pm \frac\lambda2 \Big) \prod_{a,b}^{m+n} \Gamma_h(\mu_a+\nu_b)
\;.
\eea
It has been noticed in \cite{Willett:2011gp} that this is precisely Theorem 5.5.11 of \cite{vdbult_thesis}, once specialized to parameters satisfying (\ref{condition on nu mu}), namely $\sum_a \mu_a=\sum_b \nu_b = n_f m_A$. The factors $\Gamma_h(\mu+\nu)$ correspond to the mesons $M$, and the previous factors correspond to the singlets $T$, $\tilde{T}$.

To derive the identities associated to the new dualities for chiral SQCD, we proceed to take the limits discussed in sections \ref{subsec: dual of [p,0]a theory} to \ref{subsec: FT analysis of [k,l]* duals} at the level of the partition functions. Let us define
\be
c(x)\equiv  \exp \left(\frac{\pi i }{2\,\omega_1\omega_2} x\right) \, , \qquad
\quad \zeta\equiv \exp{\left(\frac{\pi i (\omega_1^2 +\omega_2^2)}{24\,\omega_1\omega_2}\right)}\, .
\ee
In particular for $\omega_1=\omega_2=i$ we have $\zeta= e^{\pi i/12}$.
In \cite{vdbult_thesis} it is shown that
\be\label{limit on Gammah}
\lim_{x\rightarrow \pm\infty} \frac{\Gamma_{h}(x)}{\zeta^{-\sign(x)}\, c \big(\sign{(x)}(x-\omega)^2 \big)} \, = \,1   \, .
\ee
Consider integrating out a matter multiplet in the fundamental or antifundamental of the gauge group by giving it a large real mass $q_0 m_0$, corresponding to a background scalar $m_0$ for some flavor symmetry $U(1)_0$ which assigns charge $q_0$ to the matter multiplet. In the localized partition function it corresponds to the limit
\be
\lim_{m_0 \rightarrow \pm \infty} \Gamma_h(q_e x_i+ q_{\alpha}m_{\alpha} + q_0 m_0 + \Delta \omega)\, ,
\ee
where $q_e=1$ for the fundamental and $q_e=-1$ for the antifundamental representation, while $q_{\alpha}$ are the charges under to the remaining flavor symmetries $U(1)_{\alpha}$, and $\Delta$ the R-charge. From (\ref{limit on Gammah}) we see that we generate the terms
\be\label{term generated upon mass deff in Z}
\zeta^{-\sign{(q_0 m_0)}}\, \exp \Big(-\sign{(q_0 m_0)}\frac{\pi i}{2}(q_ex+ q_{\alpha}m_{\alpha} +\omega (\Delta-1)+ q_0 m_0)^2 \Big) \;.
\ee
This shows the generation of Chern-Simons terms (\ref{formula for shift of CS levels}) upon integrating out the massive matter superfield (the physical meaning of the $\zeta$ term, if any, is less clear).
There are also terms depending on the parameter $m_0$ that we take to infinity.
Using the limit (\ref{limit on Gammah}), the check of the various Seiberg-like dualities starting from (\ref{aharony duality ZeZ}) is a straightforward (albeit tedious) exercise.%
\footnote{The identities we will find have been proven rigorously in Theorems 5.5.11 and 5.5.12 of \cite{vdbult_thesis}. Here we reproduce this results with its physical meaning. Moreover there is a small mistake in \cite{vdbult_thesis} that we correct in equation (\ref{equal of Z for kl case}) below.}
The terms depending on $m_0$ cancel out between the two sides of the identity (\ref{aharony duality ZeZ}), and we find new identities for our chiral SQCD-like theories.

Remark from (\ref{term generated upon mass deff in Z}) that the CS levels for the global symmetries appear in the partition function as
\be
c\Big(2 k_{RR}\, \omega^2 \,+\, 4 \sum\nolimits_{I} \, k_{IR} \, \omega m_I \,+\, 2 \sum\nolimits_{I,J}\, k_{IJ} \, m_I m_J \Big) \;,
\ee
where here $I, J$ run over the non-R global symmetries. We note that $\omega$ enters in the partition function similarly to the real masses $m_I$ associated to the non-R symmetries. A full understanding of that fact from first principle (in the localization procedure) is left for the future.


\subsection{The $[\bm{p},\bm{0}]a$ theories}
\label{subsec:ZeqZ for [k,2] theory}

To check the duality for the $[\bm{p},\bm{0}]a$ theories of section \ref{subsec: dual of [p,0]a theory}, we take the limit (\ref{limit for k2 theory}) on both side of the identity (\ref{aharony duality ZeZ}).
Consider first the electric theory. We have to consider the limit
\be
\lim_{m_0\rightarrow -\infty} \, I_{n\, [0,0]a}^m \big( \mu+\alpha; \, \nu_{b} -\alpha, \, \nu_{\alpha}-\alpha+ m_0; \, \lambda +(s_1-s_2)m_0 \big) \big|_{x\rightarrow x-\alpha}
\ee
with $\alpha = \frac{s_1-s_2}{2s_1} m_0$, and the shift $x\rightarrow x-\alpha$ in the integrand is understood. This is better written as
\be\label{limit cIk2 elec 00}
\lim_{m_0\rightarrow -\infty} I_{n\, [0,0]a}^m \big( \mu; \, \nu_{b}, \, \nu_{\alpha}+ m_0; \, \lambda + (s_1-s_2)m_0 \big) \, c\Big( n \frac{s_1-s_2}{s_1} \big( \lambda+(s_1-s_2)m_0 \big) \, m_0 \Big) \;.
\ee
We can similarly take the limit on the magnetic side of (\ref{aharony duality ZeZ}). Explicit computation shows that all factors involving $m_0$ cancel out. We should also redefine
\be
\lambda \rightarrow \lambda - (s_1-s_2)\omega +(s_1-s_2)m_A\, ,
\ee
which corresponds to the redefinitions (\ref{def Ak2 from A,B,g}) and (\ref{redef R sym}) of the axial and R-symmetries. This leaves us with the identity
\bea
\label{duality ZeZ for p0 case}
& I_{n\, [p,0]a}^m (\mu_a;\, \nu_b;\, \lambda) \,  = &\\
 & \quad \qquad\qquad\quad I_{m\, [p,0]b}^n \big( \omega-\nu_b; \omega-\mu_a; -\lambda+(s_1-s_2) \omega \big)\,  \prod_{a,b}^{s_1, s_2} \Gamma_h(\mu_a+\nu_b)\, &\\
 &\,  \,\qquad\qquad \qquad \times  \Gamma_h\Big(-\frac12(s_1+s_2)m_A +\frac12\lambda+(m+k+1)\omega \Big)\,  \zeta^{-1} & \\
&\qquad\quad\qquad\quad \times c\Big(2 k \sum^{s_1}_a \tilde{m}_a^2 + 2ks_1 m_A^2 +2 k m \omega^2 -2k \lambda m_A  - 4k(s_1+k)m_A\omega\Big)&\\
& \qquad\quad\qquad\quad \times c\Big( \big( -\frac12(s_1+s_2)m_A -\frac12\lambda + (m-k) \omega \big)^2 \Big) \;,
\eea
with $k = -\frac12(s_1-s_2)$.
This is precisely the duality we proposed for the $[\bm{p},\bm{0}]a$ theory, including the relative CS terms for the global symmetry group.

\subsection{The $[\bm{p},\bm{q}]a$ theories}
\label{subsec:ZeqZ for [k,l] theory}

To check the duality for the $[\bm{p},\bm{q}]a$ theories at the level of the partition function, we can take the limit of section \ref{subsec: FT analysis of [k,l] duals} on the equality (\ref{aharony duality ZeZ}).
One can check that all terms involving $m_0$ cancel between the electric and magnetic sides. The identity we obtain is
\bea
\label{equal of Z for kl case}
& I_{n\, [p,q]a}^m (\mu_a;\, \nu_b;\, \lambda)\,  =  &\\
&  \quad  \,  I_{m\, [p,q]b}^n \big( \omega-\nu_b; \omega-\mu_a; -\lambda+(s_1-s_2)\omega \big) \, \prod_{a,b}^{s_1, s_2} \Gamma_h(\mu_a+\nu_b)\;  &\\
&\quad \times \,\zeta^{\frac14(s_1-s_2)^2 -k^2-2}\,    c\Big( \big( k-\frac12(s_1-s_2) \big) \sum_{a}^{s_1}\tilde{m}_a^2+ \big( k+\frac12(s_1-s_2) \big) \sum_{b}^{s_2} m_b^2\, \Big)&\\
&\quad \times \, c\Big( \, \big( k(s_1+s_2)+2 s_1s_2 \big) m_A^2 +\frac{1}{2}\lambda^2  + \big( (k+m)^2+m^2 +\frac14(s_1-s_2)^2 \big) \, \omega^2 \Big)&\\
&\quad \times \, c\Big( -2(k+m)(s_1+s_2)\omega m_A +(s_1-s_2) \lambda(m_A-\omega) \Big) \;,&
\eea
giving a nice check of our proposal.

\subsection{The $[\bm{p},\bm{q}]^*a$ theories}\label{subsec:ZeqZ for [k,l]* theory}

We can play the same game for the $[\bm{p},\bm{q}]^*a$ theories, taking the limit (\ref{limit to kl* from 22}) on the duality relation (\ref{aharony duality ZeZ}). One can show that all factors of $m_0$ cancel, and that we end up with the relation
\bea\label{ZeZ for kl*}
& I_{n\, [p,q]^*a}^m (\mu_a;\, \nu_b;\, \lambda)\, =  &\\
&  \quad  \,  I_{m\, [p,q]^*b}^n (\omega-\nu_b; \omega-\mu_a; -\lambda- 2 k \omega) \, \prod_{a,b}^{s_1, s_2} \Gamma_h(\mu_a+\nu_b)\; &\\
&\quad \times \,   c \Big( 2k \sum_{a}^{s_1}\tilde{m}_a^2 + 2k s_1 m_A^2 - 2k m\omega^2 +2 \lambda(s_1m_A -m\omega) \Big) \;. &
\eea
This agrees with the duality we proposed in section \ref{subsec: FT analysis of [k,l]* duals}.

\subsection{Complex masses and dual Higgsing}\label{subsec: Higgsing and susy breaking from Z}

In this section we consider complex mass deformations of the electric theories from the perspective of their localized partition functions. We will see how the dual description in terms of Higgsing of the magnetic gauge group arises due to distributional identities satisfied by hyperbolic gamma functions. By giving complex mass to a suitable number of flavor pairs we can Higgs completely the magnetic gauge group, reaching $m=0$; proceeding to $m<0$ brings us out of the conformal window, to a theory that either has a deformed moduli space (if $k=0$ and $s_1=s_2=n_f=n-1$, giving $m=-1$) or, most often, lacks supersymmetric vacua.

It has been observed previously \cite{Kapustin:2010mh} in the context of Giveon-Kutasov dualities that integrals that define the localized $S^3$ partition functions of superconformal theories vanish when the field theory breaks supersymmetry. A similar observation applies more generally to the Seiberg-like dualities that we are considering: $I^m_n$ integrals vanish whenever $m<0$, including the particular case when the theory is believed to have a deformed moduli space. The physical significance of this mathematical fact is however dubious: the partition functions are localized using the superconformal algebra, so that we can trust the matrix model integrals as partition functions of the field theory only if the IR superconformal algebra is visible in the UV. For instance, the partition function of the field theory on its deformed moduli space, being a free theory, does not vanish, although we find that the $I^{-1}_{1\,[0,0]}$ integral does.%
\footnote{(Added in v2:) Recently \cite{Morita:2011cs} conjectured that spontaneous supersymmetry breaking implies that the localized $S^3$ partition function $Z^{\text{loc}}$ vanishes. More precisely, they provided suggestive arguments to the effect that if the vacuum is not invariant under the supercharge $Q$ used in the localization, then $Z^{\text{loc}}$ must vanish. We thank the authors of \cite{Morita:2011cs} for discussion on this point.}

We start with the case $m=0$, where the identities (\ref{duality ZeZ for p0 case}), (\ref{equal of Z for kl case}) or (\ref{ZeZ for kl*})  provide an explicit formula for the partition function (\ref{def Zelec in 4 cases}), with the understanding that integrals on a vanishing gauge group $I_{0,\dots}^{\dots}$ are to be replaced by $1$. Starting from the partition function with $m=0$, we can obtain the partition function for $m<0$ (keeping $n$, $p$ and $q$ fixed) by setting $\mu+\nu =2\omega$ for a number of pairs $(\tilde{Q}, Q)$; as explained in Appendix \ref{App: subsec: complex mass}, this corresponds to a complex mass deformation $W=  \tilde{Q}Q$ of the electric theory. Using the identities (\ref{duality ZeZ for p0 case}), (\ref{equal of Z for kl case}) or (\ref{ZeZ for kl*}), we see that the partition function vanishes because
\be
\Gamma_h(\mu+\nu)= \Gamma_h(2\omega)=0\,
\ee
appears on the magnetic side.

On the contrary, taking $\mu_{s_1} +\nu_{s_2} = 2\omega$ for a pair $(\tilde{Q}^{s_1}, Q_{s_2})$ to reduce to $m-1$ does not lead to a vanishing result when $m>0$: the $\Gamma_h(2\omega)$ factor in the magnetic partition function is accompanied by a term which blows up in the integral over the magnetic gauge group, from $\Gamma_h(-x)\Gamma_h(x)$ in the integrand. We should rather set $\mu_{s_1} +\nu_{s_2} = 2\omega -2i\epsilon$ and then take the $\eps\to 0$ limit on both sides of the identities relating electric and magnetic partition functions.
Using the pole structure of $\Gh(z) $, one can show that
\begin{equation}\label{identity_delta_multiple}
\lim_{\epsilon\to 0} \Gh(2\omega-2i\epsilon)\,\prod_{j=1}^m \Gh(i\epsilon\pm y_j)  = \sqrt{-\omega_1\omega_2} \, \sum_{j=1}^m \delta(y_j) \prod_{i\neq j}^m \Gh(\pm y_i)\;
\end{equation}
\emph{distributionally}, \emph{i.e.} up to addition of functions with measure zero support that do not affect integration over $\int d^m y$.%
\footnote{The case $m=1$ is just the statement of duality between $N_f=1$ SQED and the XYZ model.}
The identity \eqref{identity_delta_multiple} explains, from the partition function point of view, the Higgsing pattern $U(m) \rightarrow U(m-1)$ on the magnetic side which is dual to the complex mass for a pair of fields on the electric side, allowing to reduce from ranks $(n,m)$ to ranks $(n,m-1)$ while keeping $[\bm{p},\bm{q}]$ fixed.


\section{Remarks on convergence and Z-minimization}
\label{sec: Z-min}

At the IR fixed point the superconformal R-charge is some linear combination
\be\label{R superconf trial}
R_{\text{supconf}}= R_0 + r_Q A + \Delta_M M\, ,
\ee
where $R_0$ is the trial R-charge defined above (it is either $R_{[\bm{p},\bm{q}]}$ or $R_{[\bm{p},\bm{q}]^*}$), which gives $R_0[Q]= 0$ to the quarks and antiquarks, fixing the unphysical mixing with the gauge symmetry. We have $R_{\text{supconf}}[Q]= r_Q$, while $\Delta_M$ denotes the mixing of the R-current with topological current, and it will only affect the R-charge of monopole operators (in the electric theory).  The coefficient $r_Q$ and $\Delta_M$ can be found by Z-minimization \cite{Jafferis:2010un}.

An interesting special case is when either $s_1$ or $s_2$ vanishes, so that there are no mesonic operators and no Higgs branch. The axial symmetry $U(1)_A$ is part of the gauge group and we can take $r_Q=0$ in (\ref{R superconf trial}), since the R-charge of any gauge invariant operator is independent of $r_Q$.

Let us consider the partition functions in the conformal case, when all real masses are turned off. This means we take
\be
\tilde{m}_a=m_b=0\, , \qquad m_A= \omega\, r_Q\, , \qquad \lambda =  \omega\, 2 \Delta_M\, .
\ee
in the various formulae. Moreover we will focus on the round $S^3$, so $\omega= i$. For the electric partition function,
\be\label{CFT Zelec}
Z^{\text{elec}}= Z_{U(n)_k}^{(s_1,\,s_2)}(0;0;ir_Q;i\Delta_M) \;. 
\ee
In the limit $x\rightarrow \pm \infty$, the integrand of (\ref{CFT Zelec}) takes the form
\be\label{gaussian term in limit large x}
\exp{\big( - \pi i k_{\pm} x^2 + 2\pi (\Delta_M\mp \alpha_{r_Q})x \big)}\, ,\quad\quad \text{with}\qquad \, \alpha_{r_Q}\equiv \frac{s_1+s_2}{2}(1- r_Q) - n +1\;.
\ee
Therefore the electric partition function (\ref{CFT Zelec}) converges \textit{absolutely} if and only if
\be\label{abso converg of Zelec}
0 < |\Delta_M| < \alpha_{r_Q} \, .
\ee
Consider next the magnetic dual theory. We have
\be\label{CFT Zmagn}
Z^{\text{magn}} = Z_{U(m)_{-k}}^{s_2,\,s_1} \Big( 0;0;i(1-r_Q); i \big(\frac{p-q}{2}-\Delta_M \big) \Big)\times (\text{singlets})\times (\text{phase})\, ,
\ee
where the singlets and extra phase can be read from the identities of the previous section, and $p$, $q$ in \eqref{k+-}-\eqref{The chiral electric theories}. One can check that for all cases the condition for the integral in $Z^{\text{magn}}$ to converge absolutely is that
\be
\alpha_{r_Q} + p-2 < \Delta_M < -\alpha_{r_Q} - q+2\, ,
\ee
We see that generically, for some given trial parameters $(r_Q, \Delta_M)$ the electric and magnetic partitions functions cannot both converge absolutely. In particular a necessary condition for it to happen is that $p<2$ and $q<2$.

In general the physical meaning of absolute convergence is not obvious, although it is of great help both to make sense of the formal limits of section \ref{sec: partitions functions and dualities}, allowing to commute limit and integration, and for numerical evaluations. On physical ground simple convergence is enough. When $k_{\pm}\neq 0$ in (\ref{gaussian term in limit large x}) the integral converges thanks to the Gaussian term.%
\footnote{Generally the integration contour should be taken slightly away from the real axis.}
When the effective Chern-Simons levels vanish, corresponding to the case $[\bm{p},\bm{0}]a$ with $p\geq 0$, the absolute convergence criterion (\ref{abso converg of Zelec}) is certainly physical, and corresponds to the gauge invariant monopole operator $T$ having positive dimension. This was noticed in \cite{Willett:2011gp} for the  $[\bm{0},\bm{0}]$ case.

We performed Z-minimization numerically for a few simple cases. The main new ingredient here is that we can have $\Delta_M \neq 0$ in (\ref{R superconf trial}), unlike the chiral case studied in \cite{Willett:2011gp} where charge conjugation invariance sets $\Delta_M=0$.

We also checked in several examples that whenever two theories can be connected by a RG flow, the ``free energy'' defined as
\be
F \equiv - \ln |Z|
\ee
is smaller in the IR theory. This corroborates the conjectured ``F-theorem''  \cite{Jafferis:2011zi}.

\begin{table}
\begin{center}
Case $n=1$, $k=0$\\
\begin{tabular}{| c | c |c| c| c| c |}
\hline
$s_1 \backslash s_2$ & $0$ & $1$ & $2$ & $3$ & $4$\\
\hline
\multirow{3}{*}{$0$}  & $r_Q$ &          & $r$ &       &   $r$ \\
                      & $\Delta_M$ &    -      & $0$ &   -    &   $0$ \\
                     & $-\ln|Z|$ &          & $0.9687$ &       &   $1.9901$\\
\hline
\multirow{3}{*}{$1$}  &  &    $1/3$      &  &   $0.4151$    &   \\
                      & - &    $0$      & - &   $0$  &   - \\
                     &    &    $0.8724 $ & &    $1.9548 $  &   \\
\hline
\multirow{3}{*}{$2$}  & $r$ &          & $0.4085 $&       &  $ 0.4387$ \\
                      & $0$ &    -      & $0$       &   -    &  $ 0$ \\
                     & $0.9687 $&      & $1.9340$ &       &   $2.8447$\\
\hline
\multirow{3}{*}{$3$}  &  &    $0.4151$      &  &   $0.4370 $   &   \\
                      & - &    $0 $     & - &   $0$  &   - \\
                     &    &   $ 1.9548 $ & &   $ 2.8380$   &   \\
\hline
\multirow{3}{*}{$4$}  & $r$ &          & $0.4387$ &       &   $0.4519$ \\
                      & $0$ &    -      & $0$ &   -    &   $0$ \\
                     & $1.9901$ &          & $2.8447$ &       &  $3.6791 $\\
\hline
\end{tabular}
\end{center}
\caption{Superconformal R-charges and free energy for the case $n=1$, $k=0$, and for various values of $s_1$, $s_2$. }\label{Table neq1 keq0}
\end{table}

\paragraph{Abelian theory at CS level $\bm{k=0}$.}
Consider a $U(1)$ theory with $k=0$, and generic $s_1$, $s_2$ such that $s_1+s_2$ is even. It is the simplest example of a $[\bm{p}, \bm{p}]^*$ theory, with $p=q= |s_1-s_2|$. The values of $r_Q$ and $\Delta_M$ are easily found numerically, and are reported in table \ref{Table neq1 keq0}.  In this case we still have parity symmetry, which imposes $\Delta_M=0$. The diagonal entries in table \ref{Table neq1 keq0} correspond to the $[\bm{0}, \bm{0}]$ case of section \ref{sec: aharony dual}, and the same numerical result was reported in \cite{Willett:2011gp}.
We remark that for $(s_1, s_2)=(s_1,0)$ or $(0,s_2)$, the quantity $|Z|$ is minimized for any value of $r_Q$, as expected, since in this case $r_Q$ is a mixing of the R-current with the gauge current.

The theory with $(s_1,s_2)=(2,0)$ is one of the few examples where both the electric and magnetic partition functions converge when we take $x$ real. The identity (\ref{ZeZ for kl*}) becomes
\be\nn
I_{1\, [1,1]^*a}^1(i r_Q, i r_Q, i 2 \Delta_M) \,=\, I_{1\, [1,1]^*b}^1 \big( i (1-r_Q), i(1- r_Q), -i 2 \Delta_M \big) \; e^{2\pi i (2\Delta_M r_Q-\Delta_M)}\, , \ee
which converges absolutely for $|\Delta_M|< \text{min}(r_Q, 1-r_Q)$.

\begin{table}[t]
\begin{center}
Case $n=1$, $k=-\frac12$\\
\begin{tabular}{| c | c |c| c| c| c | c|}
\hline
$s_1 \backslash s_2$ & $0$ & $1$ & $2$ & $3$ & $4$& $5$\\
\hline
\multirow{3}{*}{$0$}  & $r_Q$ &          $r$    & &          $r$    &                      & $r$  \\
                      & $\Delta_M$ &    $-\frac12 r$  & -&   $-\frac12 r+0.1404$    &   -   & $-\frac12 r+0.1841$ \\
                     & $-\ln|Z|$ &   $0.3466$   &  &   $1.5126$    &                       & $2.4459$ \\
\hline
\multirow{3}{*}{$1$}  & $r$ &          & $0.3919 $&       &  $ 0.4320$ &\\
                      & $\frac12 r$ &    -      & $-0.0437$       &   -    &  $ -0.0337$ & -\\
                     & $0.3466$&      & $1.4666$ &       &   $2.4228$ &\\
\hline
\multirow{3}{*}{$2$}  &  &    $0.3919$   &  &     $0.4278$    &     & $0.4479$\\
                      & - &    $0.0347$  & - &    $-0.0168$  &   - & $-0.0192$\\
                     &    &    $1.4666$ &    &    $2.4064$  &      & $3.2752$ \\
\hline
\multirow{3}{*}{$3$}  & $r$ &          & $0.4278$ &       &   $0.4463$ &\\
                      & $\frac12 r-0.1404$ &    -      & $0.0168$ &   -    &   $-0.0079$ &-\\
                     & $1.5126$ &          & $2.4064$ &       &  $3.2681 $&\\
\hline
\multirow{3}{*}{$4$}  &  &    $0.4320$      &  &   $0.4463 $   &   &$0.4574$\\
                      & - &    $0.0337 $     & - &   $0.0079$  &   - &$-0.0044$\\
                     &    &   $ 2.4228 $ & &   $ 3.2681$   &   &  $4.0882$ \\
\hline
\multirow{3}{*}{$5$}  & $r$ &                           & $0.4479$ &       &   $0.4574$ &\\
                      & $\frac12 r- 0.1841$ &    -      & $0.0192$ &   -    &   $0.0044$ &-\\
                     & $2.4459$ &                       & $3.2752$ &       &  $4.0882 $&\\
\hline
\end{tabular}
\end{center}
\caption{Superconformal R-charges and free energy for the case $n=1$, $k=-\frac12$, and for various values of $s_1$, $s_2$. In these theories $\Delta_M \neq 0$.}\label{Table neq1 keq-1/2}
\end{table}

\paragraph{Abelian theory at CS level $\bm{k=-\frac12}$.} Let us consider next the $U(1)$ theory at Chern-Simons level $-\frac{1}{2}$, with $s_1+s_2$ odd. In this case there is no parity symmetry and we have $\Delta_M \neq 0$. The superconformal values of $r_Q$ and $\Delta_M$ are reported in table \ref{Table neq1 keq-1/2}. In the case $s_1s_2=0$,  the partition function is minimized for any value of $r_Q$, while there is a physically meaningful contribution to $\Delta_M$.

\paragraph{$\bm{U(1)_{-1/2}}$ with a single $\bm{\tilde{Q}}$.}
The special case $U(1)$ at level $-\frac12$ with a single chiral multiplet of charge $-1$ is worth some further attention (it falls into the case $[\bm{1}, \bm{0}]a$). The flavor group is reduced to $U(1)_M \times U(1)_R$, the $U(1)_A$ symmetry being gauged.
The Seiberg dual theory is simply a free field $T$ of charge $1$ under $U(1)_M$, corresponding to the monopole of the electric gauge group.%
\footnote{See \cite{Dorey:1999rb} for an early appearance of this duality in the context of mirror symmetry.}
 Since $R[T]=\frac12$ in the IR, we must have $\Delta_M= \frac12 r_Q$, and we can set $r_Q=0$. In the magnetic theory we also have the CS levels
\be\label{kFlavor for U1 with one flavor}
k_{MM}= \frac12\, , \qquad k_{RR}= \frac18  \, , \qquad k_{MR}= -\frac14 \, .
\ee
As a function of the FI parameter $\xi$ and of the undetermined (and physically meaningless)  $r_Q$, the electric and magnetic partition functions read
\be
Z^{\text{elec}}(\xi; r_Q)\,=\, \int dx\,  e^{\frac{\pi i}{2} x^2-2\pi i \xi x + \pi r_Q x} \, \Gamma_h(-x+ i r_Q) \, ,
\ee
\be
Z^{\text{magn}}(\xi; r_Q)\,=\, \Gamma_h \Big( \xi +\frac{i}{2} \Big) \, e^{-\frac{\pi i }{12}}\, e^{-\frac{\pi i}{2}(\xi^2 -\frac{1}{4}-r_Q^2 )} \, e^{\pi \xi(r_Q-\frac{1}{2})}\, .
\ee
They are equal by (\ref{duality ZeZ for p0 case}); it is also easily checked numerically.


\section{Dualities for $Usp(2n)$ theories with fundamentals}
\label{sec: Usp}

Seiberg-like dualities for $Usp(2n)$ Yang-Mills theories with an even number of fundamentals were conjectured early on by Aharony \cite{Aharony:1997gp} and recently generalized to Chern-Simons theories by Willett and Yaakov \cite{Willett:2011gp}. For completeness, in this section we briefly review those dualities and their checks at the level of partition functions \cite{Willett:2011gp}, listing explicitly the global CS terms needed to fully specify the duality maps. Later the results of this and the next section will be used to check the $\mathcal{N}=5$ $Usp\times O$ dualities proposed in \cite{Aharony:2008gk} on the ground of a brane construction.

\subsection{Yang-Mills theories}
\label{subsec:_Usp_Ofer}

A duality for $Usp(2n)$ theory with $s=2n_f$ fundamental flavors $Q$ and no CS interactions was proposed by Aharony in \cite{Aharony:1997gp} for $n_f\geq n+1$.%
\footnote{See \cite{Karch:1997ux} for the analysis of the moduli space for any $n_f$.}
The global symmetry is $SU(2n_f)\times U(1)_A\times U(1)_R$.
For the purpose of defining the partition function of the field theory on  $S^3_b$, it turns useful to work with the Cartan subgroup $U(1)_F^{2n_f}\in U(2n_f)_F$ of the flavor group. Similarly, we will mostly work with an R-symmetry mixed with these Cartan generators. We define $m$ such that $n_f\equiv n+m+1$. Then the $Usp(2n)_0$ \textbf{electric} theory has matter content ($i,j=1\dots,n$ are gauge indices; $r,u,v=1,\dots,2n_f$ are flavor indices)\begin{equation}\label{electric_Usp_Ofer}
\begin{array}{c|c|cc|c}
& x_i & M_r & \omega & \\
 & [U(1)_i] & U(1)_{F_r} & U(1)_R & \mathrm{Mass}  \\
\hline
Q^j_u & \pm \delta_{ij} & \delta_{ur} & \Delta_u &  \pm x_j + \mu_u \\ \hline
M_{uv} & 0 & \delta_{ur}+\delta_{vr} & \Delta_u+\Delta_v & \mu_u+\mu_v\\
Y & 0 & -1 & -2n-\sum_r(\Delta_r-1) & -\sum_r \mu_r+2(m+1)\omega
\end{array}
\end{equation}
with
\begin{equation}\label{mu_r}
\mu_u\equiv M_u+\Delta_u\omega\;.
\end{equation}
$x_i$ are the scalars in the dynamical gauge vector multiplet, $M_r$ are the scalars in the background vector multiplets for the flavor symmetry, and $\omega$ is the corresponding parameter for the R-symmetry. In the last column we listed the complexifications of the real masses of the various fields, which enter the partition function.
The fields listed after the last horizontal line are composite and will not appear in the partition function of the electric theory.
The meson $M_{uv}=Q_u \cdot Q_v\equiv J_{ij}Q^i_u Q^j_v$ transforms in the 2-index antisymmetric representation of $SU(2n_f)$ due to the contraction with the antisymmetric invariant tensor $J_{ij}$ of $Usp(2n)$. The monopole operator $Y=\prod_{i=1}^n Y_i$ parametrizes the complex dimension of the Coulomb branch which is not lifted by instanton corrections. It is given in terms of the fundamental monopole operators $Y_i\sim e^{\alpha^\vee_i\cdot\,\Phi}$ (the latter expression is valid in the weak coupling region) associated to the simple coroots $\alpha^\vee_i$ of the gauge Lie algebra.

Aharony conjectured that for $n_f\geq n+1$ the electric theory is dual to a \textbf{magnetic} $Usp(2m)$ gauge theory, with $m=n_f-n-1$, with $2n_f$ fundamental flavors $q$ in the antifundamental representation of the flavor group, no CS interactions, a gauge singlet $M$  dual to the flavor antisymmetric meson, and a gauge singlet $Y$ dual to the monopole operator parametrising the Coulomb branch of the electric theory. Its matter content is
\begin{equation}\label{magnetic_Usp_Ofer}
\begin{array}{c|c|cc|c}
& y_i & M_r & \omega & \\
 & [U(1)_i] & U(1)_{F_r} & U(1)_R & \mathrm{Mass}  \\
\hline
q^{ju} & \pm \delta_{ij} & -\delta_{ur} & 1-\Delta_u &  \pm y_j +\omega- \mu_u  \\
M_{uv} & 0 & \delta_{ur}+\delta_{vr} & \Delta_u+\Delta_v & \mu_u+\mu_v\\
Y & 0 & -1 & -2n-\sum_r(\Delta_r-1) & -\sum_r \mu_r+2(m+1)\omega \\ \hline
y & 0 & +1 & -2m+\sum_r\Delta_r & \sum_r \mu_r-2m\omega
\end{array}
\end{equation}
$y$ is the monopole operator of the magnetic theory, defined similarly to $Y$, not to be confused with the scalars $y_i$ in the dynamical vector multiplet of the magnetic gauge group. Finally, a superpotential
\begin{equation}\label{W_mag}
W_{mag} = M_{rs} q^r\cdot q^s+Y y
\end{equation}
preserving all the symmetries trivializes the monopole operator $y$ and the meson of magnetic quarks in the chiral ring.

To test the conjectured duality, we introduce the partition function on $S^3_b$ of a symplectic gauge theory with fundamentals. We define the integrals \cite{vdbult_thesis}
\begin{equation}\label{J_Usp}
J_{n,\,s,\,t}(\mu)\equiv \frac{1}{\sqrt{-\omega_1\omega_2}^n 2^n n!} \int_{C^n} \prod\limits_{1\leq j<k\leq n} \frac1{\Gh(\pm x_j\pm x_k)} \prod_{j=1}^n  \frac{\prod\limits_{r=1}^s \Gh(\mu_r\pm x_j)}{\Gh(\pm 2x_j)} \, c(2t x_j^2) \, dx_j \;,
\end{equation}
with $t+s\in 2\bZ$. $2^n n!$ is the order of the Weyl group of $Usp(2n)$.
Then the partition of the $Usp(2n)_k$ gauge theory with $s$ fundamentals with mass parameters $\mu_r$ associated to the global symmetry group is
\begin{equation}\label{partition_func_phys}
Z_{Usp(2n)_k}^s (\mu) = J_{n,\,s,\,2k}(\mu)\;, \qquad \mathrm{where}\qquad \mu_r= M_r+ \Delta_r \omega\;.
\end{equation}
$t+s\in 2\bZ$ translates into the quantization of the CS level $k+\frac{s}{2}\in\bZ$ which ensures invariance of the partition function under large gauge transformations.%
\footnote{Note that a CS term at level $k$ for $Usp(2n)$ contributes $\exp(-2\pi i k \sum_{j=1}^n x_j^2 )=c(4k \sum_{j=1}^n x_j^2)$. The extra factor of $2$ compared to the $U(n)$ case is due to normalization of the generators \cite{Willett:2011gp}.}

As pointed out in \cite{Willett:2011gp}, the proof of the duality of \cite{Aharony:1997gp} at the level of partition functions is provided by the transformation identity Theorem 5.5.9 of \cite{vdbult_thesis}, which we rewrite as
\begin{equation}\label{duality_Usp_Ofer}
Z_{Usp(2n)_0}^{2n_f} (\mu) = Z_{Usp(2m)_{0}}^{2n_f}(\omega-\mu)\;\; \Gh \Big( 2(m+1)\omega-\sum_{r=1}^{2n_f}\mu_r \Big) \prod_{1\leq r<s\leq 2n_f} \Gh(\mu_r+\mu_s)\;,
\end{equation}
with $m=n_f-n-1\geq 0$. The LHS is the partition function of the electric theory \eqref{electric_Usp_Ofer}. The RHS is the partition function of the magnetic theory \eqref{magnetic_Usp_Ofer}: the first piece accounts for the $Usp(2m)$ theory with $2n_f=2(n+m+1)$ fundamentals, the second piece for the singlet $Y$ dual to the monopole, and the third piece for the antisymmetric mesons $M$.


\subsection{Chern-Simons theories}

Willett and Yaakov derived dualities for $Usp(2n)$ Chern-Simons theories with fundamental flavors, by means of real mass deformations of Aharony's dual theories \cite{Willett:2011gp}. Here we repeat their computation and supplement it with the global Chern-Simons couplings induced by the massive fermions.
The relevant identity that we will need is
\begin{equation}
Z_{Usp(2n)_k}^s(\mu) = \lim_{m_0\to\sign(k)\infty} \zeta^{4nk} \, c\big( -4nk(m_0-\omega)^2 \big) \, Z_{Usp(2n)_0}^{s+2|k|}(\mu,\underbrace{m_0,\dots,m_0}_{2|k|})\;.
\end{equation}
In physical terms, we are giving infinite real mass (with same sign as $k$) to $2|k|\in\bZ$ quarks of the ultraviolet YM theory, thus producing CS interactions at level $k$ for the IR gauge theory.
Applying this limit to both sides of Aharony's duality \eqref{duality_Usp_Ofer}, we find the identity
\begin{equation}\label{duality_Usp_CS}
\begin{split}
&Z_{Usp(2n)_k}^{s} (\mu) = Z_{Usp(2m)_{-k}}^{s}(\omega-\mu) \prod_{1\leq u<v\leq s} \Gh(\mu_u+\mu_v)\,  \zeta^{\sign(k)(2|k|-1)(|k|-1)} \\
&\times c \Big( -4nk\omega^2 - \sign(k) \Big[ (2m+1)\omega - \sum_{r=1}^s\mu_r \Big]^2 + 2k\sum_{r=1}^s(\mu_r-\omega)^2+k(2|k|-1)\omega^2 \Big)\;,\\
& \quad m = |k|+\frac{s}{2}-n-1\geq 0\;,\quad k\neq 0\;.
\end{split}
\end{equation}
The sum inside the last term encodes the global CS terms due to the massive fields: massive electric quarks (this contribution is moved from the LHS to the RHS), $Y$, mesons $M_{u,\,s+a}$, and mesons $M_{s+a,\,s+b}$ respectively.

The identity \eqref{duality_Usp_CS} reflects the duality of \cite{Willett:2011gp} between the \textbf{electric}
$Usp(2n)_k$ theory with $s$ fundamentals $Q$ (with $k+\frac{s}{2}\in\bZ$),
\begin{equation}\label{electric_Usp_CS}
\begin{array}{c|c|cc|c}
& x_i & M_r & \omega & \\
 & [U(1)_i] & U(1)_{F_r} & U(1)_R & \mathrm{Mass} \\
\hline
Q^j_u & \pm \delta_{ij} & \delta_{ur} & \Delta_u &  \pm x_j + \mu_u \\ \hline
M_{uv} & 0 & \delta_{ur}+\delta_{vr} & \Delta_u+\Delta_v & \mu_u+\mu_v
\end{array}
\end{equation}
and vanishing superpotential, and the \textbf{magnetic} $Usp(2m)_k$ theory with $s$ fundamentals $q$ and singlets $M_{uv}$ dual to the electric mesons,
\begin{equation}\label{magnetic_Usp_CS}
\begin{array}{c|c|cc|c}
& y_i & M_r & \omega & \\
 & [U(1)_i] & U(1)_{F_r} & U(1)_R & \mathrm{Mass}  \\
\hline
q^{ju} & \pm \delta_{ij} & -\delta_{ur} & 1-\Delta_u &  \pm y_j +\omega - \mu_u \\
M_{uv} & 0 & \delta_{ur}+\delta_{vr} & \Delta_u+\Delta_v & \mu_u+\mu_v
\end{array}
\end{equation}
with superpotential
\begin{equation}\label{W_mag_CS}
W_{mag} = M_{uv} q^u\cdot q^v
\end{equation}
when $m= \frac s2 + |k| -n-1\geq 0$ and $k\neq 0$. Again $\mu_r$ is given by \eqref{mu_r}.
In addition there are relative global Chern-Simons terms that can be read from \eqref{duality_Usp_CS}. For simplicity we list them here only in the case of maximal global symmetry $U(1)_A\times SU(s)\times U(1)_R$ and for an R-symmetry mixed with $A$, giving $R[Q]=\Delta$:
\bea
& \Delta k_{RR} = \frac{1}{2}\sign(k)\left[-(2m+1-s \Delta)^2+2|k| s (\Delta-1)^2 +|k|(2|k|-1-4n)\right]\\
& \Delta k_{R A} = -\sign(k) \frac{s}{2} \left[ 2n+1-s +(s-2|k|)\Delta\right]\\
& \Delta k_{A A} = -\sign(k)\frac{s}{2}(s-2|k|)\\
& \Delta k_{SU(s)} = k \;.
\eea
Notice that when $m <0$ supersymmetry is broken and the matrix integral vanishes.


\section{Dualities for $O(N_c)$ theories with vector matter}
\label{sec: O}

Seiberg-like dualities for Chern-Simons theories with orthogonal gauge groups and matter in the vector representations have been recently conjectured by Kapustin \cite{Kapustin:2011gh}.
The duality states that the \textbf{electric} $\cN=2$ $O(N_c)_k$ CS theory with $N_f$ flavors of vector matter $Q$ in the fundamental of the $U(N_f)$ flavor group%
\footnote{The Chern-Simons level $k$ is an integer irrespective of $N_f$.}
\begin{equation}\label{electric_O_CS}
\begin{array}{c|c|cc}
 & [O(N_c)] & U(N_f)  & U(1)_R   \\
\hline
Q & \bm{N_c} & \bm{N_f} & \Delta
\end{array}
\end{equation}
is dual to a \textbf{magnetic} $O(\tilde{N}_c)_{-k}$ theory with $N_f$ flavors of vector matter $q$ in the antifundamental of $U(N_f)$ and a gauge singlet $M$ in the two-index symmetric representation of $U(N_f)$ (dual to the meson of electric quarks)
\begin{equation}\label{magnetic_O_CS}
\begin{array}{c|c|cc}
 & [O(\tilde{N}_c)] & U(N_f) &  U(1)_R  \\
\hline
q & \bm{\tilde{N}_c} & \bm{\overline{N_f}} & 1-\Delta \\
M & \bm{1} & \bm{N_f(N_f+1)/2} & 2\Delta
\end{array}
\end{equation}
and a superpotential
\begin{equation}\label{W_mag_CS_O}
W_{mag} = M_{uv} \, q^u\cdot q^v
\end{equation}
when $\tilde{N}_c=N_f+|k|+2-N_c\geq 0$. The choice of $O$ rather than $SO$ gauge groups is crucial for the match of global symmetries and moduli spaces of dual theories.
In the original paper it was checked numerically for several low rank dual pairs that the absolute values of the electric and magnetic partition functions coincide. Another successful test by means of the superconformal index between dual pairs was performed recently for some low rank examples \cite{Hwang:2011qt}. In this section we provide a proof that the partition functions of dual theories on $S^3_b$ match, once certain global CS terms are included, and extend the duality to the IR fixed points of orthogonal Yang-Mills theories.

When writing the partition function we have to distinguish between $N_c$ even and odd, due to the different Lie algebras. The partition function of a $O(2n)_k$ theory with $N_f$ flavors of vectors is
\begin{equation}\label{Z_O_2n}
Z_{O(2n)_k}^{N_f}(\mu)\equiv \frac{1}{2^{n-1} n!} \int \prod\limits_{1\leq j<k\leq n} \frac1{\Gh(\pm x_j\pm x_k)} \prod_{j=1}^n  \prod_{r=1}^{N_f} \Gh(\pm x_j+\mu_r)\, c(2 k x_j^2) dx_j \;,
\end{equation}
where $\mu_r=M_r+\Delta_r \omega$, working again with the Cartan subalgebra of the global symmetry group. The partition function of a $O(2n+1)_k$ theory with $N_f$ flavors is
\begin{equation}\label{Z_O_2n+1}
Z_{O(2n+1)_k}^{N_f}(\mu)\equiv \frac{1}{2^n n!} \prod_{r=1}^{N_f} \Gh(\mu_r) \int \prod\limits_{1\leq j<k\leq n} \frac1{\Gh(\pm x_j\pm x_k)}\prod_{j=1}^n  \frac{\prod\limits_{r=1}^{N_f} \Gh(\pm x_j+\mu_r)}{\Gh(\pm x_j)}  \, c(2 k x_j^2) dx_j \;.
\end{equation}
By means of a trick analogous to that used in \cite{Dolan:2008qi} for  the superconformal indices of 4d gauge theories, we can reduce the partition functions of the orthogonal gauge theories to those of symplectic gauge theories with fundamental matter \eqref{partition_func_phys}-\eqref{J_Usp} (see also \cite{Spiridonov:2011hf}).
We will exploit the identity  \cite{Kurokawa_dS}
\begin{equation}\label{identity_Gamma_h}
\Gh(2z)=\Gh(z) \Gh \Big( z+\frac{\omega_1}{2} \Big) \Gh\Big( z+\frac{\omega_2}{2} \Big) \Gh(z+\omega)\;.
\end{equation}
Defining
\begin{equation}\label{U_V}
\begin{split}
& U= \Big( 0,\frac{\omega_1}{2},\frac{\omega_2}{2},\omega\Big)\qquad \qquad V=\Big(0,\frac{\omega_1}{2},\frac{\omega_2}{2}\Big) \\
& U'=\Big(\frac{\omega_1}{2},\frac{\omega_2}{2} \Big)\qquad \qquad\quad\;\;\; V' = \Big(\frac{\omega_1}{2},\frac{\omega_2}{2},\omega \Big)
\end{split}
\end{equation}
and applying formulas \eqref{reflexion formula for Gh} and \eqref{special_values}, we get \begin{align}
&\Gh(2z)=\prod_{a=1}^4 \Gh(z+U_a)&  &\Gh(\pm 2z)=\prod_{a=1}^3 \Gh(\pm z+V_a) \\
&\frac{\Gh(2z)}{\Gh(z)}=\prod_{a=1}^3 \Gh(z+V'_a)&  & \frac{\Gh(\pm 2z)}{\Gh(\pm z)}=\prod_{a=1}^2 \Gh(\pm z+U'_a)\\
& \prod_{a<b}^4 \Gh(U_a+U_b) =1 &  & \prod_{a<b}^3 \Gh(V_a+V_b)=\frac{1}{2}\\
& \prod_{a<b}^3 \Gh(V'_a+V'_b)=2 &  & \prod_{a<b}^2 \Gh(U'_a+U'_b)=1\;.
\end{align}
We use these identities to rewrite the partition functions of the orthogonal theories in term of those of the symplectic theories. If $N_c=2n$ is even,
\begin{equation}
Z_{O(2n)_k}^{N_f}(\mu) =
\begin{cases}
2\, Z_{Usp(2n)_{k/2}}^{N_f+4}(\mu,U) \;,\quad  & N_f+k \in 2\bZ \\
2\, Z_{Usp(2n)_{k/2}}^{N_f+3}(\mu,V) \;,\quad  & N_f+k \in 2\bZ+1
\end{cases}
\end{equation}
and if $N_c=2n+1$ is odd
\begin{equation}
Z_{O(2n+1)_k}^{N_f}(\mu) =
\begin{cases}
 \prod\limits_{r=1}^{N_f}\Gh(\mu_r)\,\cdot \, Z_{Usp(2n)_{k/2}}^{N_f+2}(\mu,U') \;,\quad  & N_f+k \in 2\bZ \\
 \prod\limits_{r=1}^{N_f}\Gh(\mu_r)\,\cdot\, Z_{Usp(2n)_{k/2}}^{N_f+3}(\mu,V') \;,\quad & N_f+k \in 2\bZ+1
\end{cases}
\end{equation}
The identities encoding Kapustin's duality for partition functions on the squashed $S^3_b$ then follow from those for Willett-Yaakov's duality \eqref{duality_Usp_CS}.

We start by considering $k=0$, corresponding to the IR fixed point of Yang-Mills theory with $N_f$ flavors of matter in the vector representation, and use the identity \eqref{duality_Usp_Ofer} for Aharony's duality of symplectic SQCD theories. The four cases with $N_c$ and $N_f+k$ even and odd boil down to the single formula
\begin{equation}\label{duality_0_YM}
Z_{O(N_c)_0}^{N_f} (\mu) = Z_{O(\tilde{N}_c)_{0}}^{N_f}(\omega-\mu)\; \Gh \Big( \tilde{N}_c \omega - \sum_{r=1}^{N_f}\mu_r \Big) \prod_{1\leq r\leq s \leq N_f} \Gh(\mu_r+\mu_s)\;,
\end{equation}
with $\tilde{N}_c=N_f+2-N_c\geq 0$.
This identity suggests that the \textbf{electric} $O(N_c)_0$ theory with $N_f$ flavors $Q$ in the fundamental of $U(N_f)$,
\begin{equation}\label{electric_O_YM}
\begin{array}{c|c|ccc}
 & [O(N_c)] & SU(N_f) & U(1)_A  & U(1)_R   \\
\hline
Q & \bm{N_c} & \bm{N_f} & 1 & \Delta
\end{array}
\end{equation}
is dual to a \textbf{magnetic} $O(\tilde{N_c})_0$ theory with $N_f$ flavors $q$ in the antifundamental of $U(N_f)$, singlets $M$ in the 2-index symmetric of $U(N_f)$ dual to the electric mesons, and a singlet $Y$ dual to a monopole operator of the electric theory with minimal magnetic flux,
\begin{equation}\label{magnetic_O_YM}
\begin{array}{c|c|ccc}
 & [O(\tilde{N}_c)] & SU(N_f)& U(1)_A &  U(1)_R  \\
\hline
q & \bm{\tilde{N}_c} & \bm{\overline{N_f}} & -1 & 1-\Delta \\
M & \bm{1} & \bm{N_f(N_f+1)/2} & 2 & 2\Delta \\
Y & \bm{1} & \bm{1} & -N_f & \tilde{N}_c - N_f\Delta
\end{array}
\end{equation}
together with a superpotential
\begin{equation}\label{W_mag_YM}
W_{mag} = M_{rs} \,q^r\cdot q^s+Y y
\end{equation}
coupling the singlet $Y$ to a monopole operator of minimal magnetic flux in the magnetic theory. 
Magnetic fluxes $B$ (in the Cartan subalgebra of the Lie algebra of the gauge group G) inserted by monopole operators are constrained by Dirac quantization $\exp(2\pi i B)=id_G$ \cite{Kapustin:2005py}. Let us introduce generators $H_j$ of the Cartan subalgebra: $H_j$ has vanishing entries except for $\bigl(\begin{smallmatrix} 0 & i\\ -i & 0 \end{smallmatrix} \bigr)$ in the $j$-th diagonal 2 by 2 block. $j$ runs from 1 to the rank $r$ of the group. The solution of the quantization condition for orthogonal groups is $B=\sum_j m_j H_j$ with integer $m_j$.%
\footnote{Note that the lattice of allowed magnetic fluxes for an orthogonal group is finer than the coroot lattice of the associated Lie algebra \cite{Kapustin:2005py}, since orthogonal groups have nontrivial fundamental group. Monopole operators with minimal flux do not lie in the coroot lattice.}
A monopole operator with minimal flux turns on a single $m_j$ with magnitude 1: such a minimal monopole operator in the electric theory has the correct quantum numbers to be dual to the singlet $Y$ in \eqref{magnetic_O_YM}; moreover the quantum numbers of the minimal monopole operator $y$ in the magnetic theory allow the superpotential term $Y y$ in \eqref{W_mag_YM}.

From the quantum numbers of the gauge invariants $M_{rs}$ and $Y$ parametrizing the moduli space that are listed in \eqref{magnetic_O_YM}, we see that the effective superpotential on the moduli space is, with a suitable normalization,
\begin{equation}\label{W_eff_O}
W_{eff} = \tilde{N_c} (Y^2 \det M)^\frac{1}{\tilde{N_c}}
\end{equation}
if $\tilde{N_c}=N_f+2-N_c\geq 1$, namely $N_f \geq N_c-1$.
As usual in Seiberg-like dualities, the singularity at the origin is cured by the introduction of the dual gauge group.
Adding a superpotential mass term allows us to flow to theories with lower $N_f$. When $N_f = N_c-1$, the magnetic gauge group is $O(1)=\bZ_2$, the effective superpotential (\ref{W_eff_O}) is regular and contributes to the exact superpotential of the magnetic description, $W_{mag}=M q^2+ Y^2 \det M$. If $N_f=N_c-2$ the dual gauge group is trivial and the theory has a deformed moduli space described by the constraint $Y^2 \det M+ q^2=1$ (the strong coupling scale has been set to unity)%
\footnote{This paragraph was corrected in v3 after \cite{Aharony:2011ci} appeared, which studies these special cases in more details.}. Finally there are no supersymmetric vacua if $\tilde{N_c}<0$, namely $N_f<N_c-2$.

The identities encoding the dualities of \cite{Kapustin:2011gh} for Chern-Simons theories then follow by real mass deformations of the duality identity \eqref{duality_0_YM} for the Yang-Mills theories. The result is
\begin{equation}\label{duality_O_CS}
\begin{split}
&Z_{O(N_c)_k}^{N_f} (\mu) = Z_{O(\tilde{N_c})_{-k}}^{N_f}(\omega-\mu) \prod_{1\leq u\leq v\leq N_f} \Gh(\mu_u+\mu_v)\,  \zeta^{\frac{1}{2}\sign(k)(|k|+1)(|k|+2)} \\
&\times c \Big( - N_c k\omega^2 -\sign{k} \Big[ (\tilde{N}_c-1)\omega-\sum_{r=1}^{N_f}\mu_r \Big]^2 + k\sum_{r=1}^{N_f} (\mu_r-\omega)^2 + \frac{k}{2}(|k|+1)\omega^2 \Big) \;,\\
& \quad \tilde{N_c} = N_f+ |k|+2 - N_c\geq 0\;,\quad k\neq 0\;,
\end{split}
\end{equation}
which agrees with Kapustin's conjecture and provides the relative global CS terms
\bea
& \Delta k_{RR} = \frac{1}{4}k (-2N_c+|k|+1) +\frac{1}{2} k N_f(\Delta-1)^2 -\frac{1}{2}\sign(k) (\tilde{N_c}-1-N_f\Delta)^2 \\
& \Delta k_{R A} = \frac{1}{2}\sign(k) N_f \left[(|k|-N_f) \Delta + \tilde{N_c}-1-|k| \right]\\
& \Delta k_{A A} = \frac{1}{2} \sign(k)N_f (|k|-N_f)\\
& \Delta k_{SU(N_f)} = \frac{k}{2} \;.
\eea

We close this section remarking that, equipped with the global CS terms for the $Usp$ and $O$ dualities, one can check at the level of partition functions the dualities for $\cN=5$ $Usp(2N_{USp})_{k}\times O(N_O)_{-2k}$ theories put forward in \cite{Aharony:2008gk} on the ground of a brane construction. The partition functions of dual theories match, up to a phase due to powers of $\zeta$ and RR global CS terms, as in the case of unitary ABJ dualities \cite{Kapustin:2010mh}.


\section{Conclusions}

In this paper we  studied Seiberg-like dualities between $\cN=2$ gauge theories in three dimensions. We proposed new such dualities for $U(n)$ Yang-Mills and Chern-Simons theories with fundamental and antifundamental matter in complex representation of the gauge group (often referred to as ``chiral'', with an abuse of terminology). All these dualities follows from the one of Aharony \cite{Aharony:1997gp}. We were careful to keep track of various relative Chern-Simons levels for the global symmetry group and we stressed that they are crucial to the duality map. We also sketched an application of our results to Seiberg dualities for Chern-Simons quivers.

Similarly, we specified which global Chern-Simons terms should be included in previously known dualities for Chern-Simons-matter theories with symplectic \cite{Willett:2011gp} and orthogonal \cite{Kapustin:2011gh} gauge groups. We also proposed an Aharony-like duality for $O(n)$ gauge theories with Yang-Mills kinetic term.

We have checked that all these dualities hold at the level of the partition function. The corresponding identities for the localized partition functions on $S^3_b$ follow from beautiful recent mathematical results \cite{vdbult_thesis}, which we checked and explained in physical terms.

There are several obvious roads to travel from here. First, it would be interesting to further check our proposals and investigate the structure of the quantum chiral ring of these theories more in depth. An important tool for that is the superconformal index \cite{Imamura:2011su, Krattenthaler:2011da}: it would be nice to check that it matches between dual theories. In particular the global Chern-Simons terms will be important when dealing with the generalized superconformal index recently proposed in \cite{Kapustin:2011jm}.

Secondly, we have left one untied knot in our characterization of the Chern-Simons levels for the R-symmetry background gauge field. To clarify the origin of the dependence of the Lagrangian on $\omega$, we should work in supergravity with an off-shell SUGRA multiplet containing $A_{\mu}^{(R)}$, and see which background supergravity fields should be turned on in order to preserve supersymmetry on $S^3_b$. This program allowed \cite{Festuccia:2011ws} to explain why $\omega$ enters like in (\ref{mixing with R and omega}) in the case of the round sphere ($\omega=i$). One should perform the same kind of analysis for the squashed $S^3$ and also explain how the $\cN=2$ completion of the CS interaction $A^{(R)}\wedge dA^{(R)}$ leads to the term $k_{RR} \omega^2$ in the field theory Lagrangian on $S^3_{b}$.

Lastly, the dualities studied here can be used as building blocks for Seiberg-like dualities in more complicated theories, such as the quiver theories which describe M2-branes at Calabi-Yau fourfold singularities. This certainly deserves further investigation \cite{to:appear}.

\bigskip

\section*{Acknowledgments}

\vspace{-5pt}
We would like to thank Ofer Aharony and Luca Ferretti for valuable discussions.
FB thanks the High Energy Group at the Weizmann Institute and the 9th Simons Summer Workshop for hospitality and a nice atmosphere. CC thanks the PMIF of ULB for hospitality.
The work of FB was supported in part by the US NSF under Grants No. PHY-0844827
and PHY-0756966. CC is a Feinberg Postdoctoral Fellow at the Weizmann Institute for
Sciences. SC was supported by the Israel Science Foundation (grant 1468/06), the German-
Israel Project Cooperation (grant DIP H52), the German-Israeli Foundation (GIF) and the
US-Israel Binational Science Foundation (BSF).


\appendix

\section{The $S^3_b$ partition function and $\Gamma_h(z)$}
\label{app: Gammah}

Let us consider an $\cN=2$ matter multiplet $\Phi$ of R-charge $\Delta$ and  charges $q_{\alpha}$ under various Abelian groups $U(1)_{\alpha}$ (which could be part of the gauge group or the flavor group, in the overall theory). The scalar component $\phi$ couples to real scalars $m_{\alpha}$ in vector multiplets (including background ones for global symmetries) through
\be
V= m^2 |\phi|^2\, , \qquad \qquad m \equiv \sum_{\alpha}\, q_{\alpha}m_{\alpha}
\ee
When we consider the partition function on $S^3$, $\Phi$ gives a contribution \cite{Kapustin:2009kz,Jafferis:2010un, Hama:2010av}
\be\label{Z chiral mult S3}
Z_{S^3}^{\Phi} = \prod_{n \geq 1}^{\infty} \left(\frac{n+1+ i m - \Delta}{n-1- i m + \Delta} \right)^n \;.
\ee
It can be generalized to the partition function for a $U(1)\times U(1)$ isometrically squashed 3-sphere (or hyperellipsoid) $S^3_b$ \cite{Hama:2011ea}:
\be\label{Z chiral mult squashed S3}
Z_{S^3_b}^{\Phi} = \prod_{n_1, n_2 \geq 0}^{\infty} \frac{(n_1+1)b +(n_2+1)b^{-1}+ i m - \frac{b+b^{-1}}{2} \Delta}{ n_1b +n_2 b^{-1}- i m + \frac{b+b^{-1}}{2} \Delta} \, ,
\ee
with $b$ the squashing parameter ($b=1$ is the round sphere). As they stand, these infinite products are divergent.
There is a very natural regularization of (\ref{Z chiral mult squashed S3}) in term of the hyperbolic gamma function introduced by Ruijsenaars \cite{Ruijsenaars}, or equivalently the inverse of the double sine function of Koyama and Kurokawa  \cite{Kurokawa_dS}. We will follow the notation of van de Bult \cite{vdbult_thesis}, and recall the relation between these various functions in \eqref{relation_vdB_Ruijs_KK}.

First we introduce the double gamma function $\Gamma_2$, defined as the analytic continuation of
\be
\Gamma_2(z; \omega_1, \omega_2) = \exp{\bigg( \partial_s \Big(\sum_{n_1,  n_2 \geq 0} (n_1 \omega_1+n_2\omega_2 + z)^{-s} \Big) \Big|_{s=0}\bigg)} \;.
\ee
Then the \textit{hyperbolic gamma function} is defined as
\be
\Gamma_h(z; \omega_1, \omega_2) \,=\, \frac{\Gamma_2(z; \omega_1, \omega_2)}{\Gamma_2(\omega_1+\omega_2 -z; \omega_1, \omega_2)}\,  ,
\ee
which formally gives
\be\label{Gammah as infinite product}
\Gamma_h(z; \omega_1, \omega_2) = \prod_{n_1,n_2\geq 0} \frac{ (n_1+1)\omega_1 +(n_2+1) \omega_2 - z}{n_1\omega_1 +n_2 \omega_2 + z}\, .
\ee
Comparing with (\ref{Z chiral mult squashed S3}), we see that
\be\label{ZS3 eq Gammah}
Z_{S^3_b}^{\Phi}(z) \, = \, \Gamma_h(z; \omega_1, \omega_2) \, , \qquad \text{with}\qquad \omega_1= ib\, , \quad \omega_2= i b^{-1}\, , \quad z= m + \omega \Delta\, ,
\ee
where we have defined
\be
\omega\equiv \frac{\omega_1+\omega_2}{2}\, .
\ee
Physically, for the squashed $S^3_b$ the parameter $\omega$ is purely imaginary with $|\omega|>1$, while $\omega=i$ for the round $S^3$.
The presentation (\ref{Gammah as infinite product}) makes it clear that $\Gamma_h(z)$ has poles at
\be
z_{\text{pole}}= -n_1\omega_1 - n_2 \omega_2\, , \qquad n_1, n_2 \in \bZ_{\geq 0}\, ,
\ee
and zeros at $z= m_1 \omega_1+ m_2 \omega_2$ for $m_1, m_2 \in \bZ_{\geq 1}$.
Poles and zeros are simple unless $\omega_1/\omega_2 \in \mathbb{Q}$.
In the physical limit (\ref{ZS3 eq Gammah}) the poles and zeros fall on the imaginary axis.

Van de Bult's hyperbolic gamma function $\Gh(z;\omega_1,\omega_2)$ is related to Ruijsenaars' hyperbolic gamma function $G(a_+,a_-;z)$ \cite{Ruijsenaars} and to
the double sine function $S_2(z;a_1,a_2)$ of \cite{Kurokawa_dS} as follows:
\begin{equation}\label{relation_vdB_Ruijs_KK}
\Gamma_h(z;\omega_1,\omega_2) = G(-i\omega_1,-i\omega_2; z-\omega)=S_2^{-1}(-iz;-i\omega_1,-i\omega_2).
\end{equation}
The hyperbolic gamma function also has the following integral representation \cite{vdbult_thesis}:
\be
\Gamma_h(z; \omega_1, \omega_2) = \exp\left(i \int_0^{\infty} \left(\frac{z-\omega}{\omega_1\omega_2 x} - \frac{\sin(2x(z-\omega))}{2\sin(\omega_1 x)\sin(\omega_2 x)} \right)\frac{dx}{x} \right)\, ,
\ee
for $z$ such that $0<\im{(z)}<\im{(\omega_1+\omega_2)}$.
It satisfies the difference equations
\bea\label{difference eqt}
\Gamma_h (z+\omega_1)&=& 2 \sin \Big( \frac{\pi z}{\omega_2} \Big) \Gamma_h(z)\, , \\
\Gamma_h (z+\omega_2)&=& 2 \sin\Big(\frac{\pi z}{\omega_1} \Big) \Gamma_h(z)\, ,
\eea
the reflection formula
\be\label{reflexion formula for Gh}
\Gamma_h (z)\Gamma_h (\omega_1+\omega_2 -z) = 1\, ,
\ee
and takes the values
\be\label{special_values}
\begin{split}
\Gh(\omega)= 1\;,\qquad &\Gh\Big( \frac{\omega_1}{2}\Big) = \Gh\Big( \frac{\omega_2}{2}\Big) = \frac{1}{\sqrt{2}}\;,\qquad \Gh\Big( \omega+\frac{\omega_1}{2}\Big)= \Gh\Big(\omega+\frac{\omega_2}{2}\Big) = \sqrt{2}\\
&\Gh(\omega_1) = \sqrt{\frac{\omega_1}{\omega_2}}\;,\qquad \Gh(\omega_2) = \sqrt{\frac{\omega_2}{\omega_1}}\;.
\end{split}
\ee
We also use the identity  \cite{Kurokawa_dS}
\begin{equation}
\Gh(2z)=\Gh(z)\Gh\Big(z+\frac{\omega_1}{2}\Big)\Gh\Big(z+\frac{\omega_2}{2} \Big)\Gh(z+\omega)\;.
\end{equation}

In this paper we always write the partition function $Z^{\Phi}_{S^3_b}$ as $\Gamma_h$, suppressing the periodicities in \eqref{ZS3 eq Gammah} from the notation:
\be
Z_{S^3_b}^{\Phi}(z) =  \Gamma_h(z)\, \qquad \qquad \text{with} \quad z= m +\omega \Delta\, .
\ee
Finally we use van de Bult's shorthand notation
\begin{equation}
\Gh(a\pm x)\equiv \Gh(a+x) \Gh(a-x) \;,
\end{equation}
so that \eqref{reflexion formula for Gh} and \eqref{special_values} imply
\begin{equation}
\frac{1}{\Gh(\pm x)}= -4 \sin \Big( \frac{\pi z}{\omega_1} \Big) \sin \Big( \frac{\pi z}{\omega_2} \Big)\:.
\end{equation}
The last identity allows us to express the contribution of vector multiplets to the $S^3_b$ partition function \cite{Hama:2011ea} in terms of hyperbolic gamma functions as well.

\subsection{$Z^{\Phi}_{S^3}$ in term of Jafferis' function}

In the case of the round $S^3$, $\omega_1=\omega_2=i$,  we have the identity \cite{Ruijsenaars,Jafferis:2010un}
\be\label{dif equ for s2}
\frac{d}{d{\tilde{z}}} \ln \Gamma_h(i(1-\tilde{z});i,i) = -\pi \tilde{z} \cot{(\pi \tilde{z})}\, .
\ee
Introducing the complex parameter $\tilde{z}=1-\Delta + i m = 1+iz$, and defining the function
\be
l(\tilde{z}) = \ln{\Gamma_h(i(1-\tilde{z});i,i)}\, ,
\ee
one can integrate (\ref{dif equ for s2}) to find
\bea\label{explicit lz}
\text{if}\quad  \im{\tilde{z}} < 0 &:&  \quad l(\tilde{z})&= \frac{1}{2\pi i} \text{Li}_2(e^{-2\pi i \tilde{z}}) -  \tilde{z} \ln(1-e^{-2\pi i \tilde{z}}) -\frac{\pi i \tilde{z}^2}{2}+\frac{\pi i}{12}\; , \\
\text{if}\quad  \im{\tilde{z}} > 0 &:& \quad l(\tilde{z})&= -\frac{1}{2\pi i} \text{Li}_2(e^{2\pi i \tilde{z}}) -  \tilde{z} \ln(1-e^{2\pi i \tilde{z}}) +\frac{\pi i \tilde{z}^2}{2}-\frac{\pi i}{12}\;.
\eea
This function $l(\tilde{z})$ was introduced by Jafferis in \cite{Jafferis:2010un}, although it appeared before in the mathematical literature \cite{Ruijsenaars,Kurokawa_dS}. We have
\be
Z_{S^3}^{\Phi}(z) =  \Gamma_h(z; i; i) = e^{l(1+ i z)}\, ,
\ee
which is perhaps the most useful expression for explicit numerical computations.

\subsection{Complex masses and integrating out}
\label{App: subsec: complex mass}

Consider two chiral superfields $\Phi$ and $\tilde{\Phi}$ which are in conjugate representation of the gauge and flavor groups. They contribute
\be
\Gamma_h(m +\omega \Delta)\, \Gamma_h(-m +\omega \tilde{\Delta})\,
\ee
to the integrand of the partition function. Now consider adding a superpotential term
\be
W= \mu \Phi \tilde{\Phi}\, .
\ee
This gives a constraint $\Delta+\tilde{\Delta}=2$ on the R-charges, and
\be
\Gamma_h(m +\omega \Delta)\,\Gamma_h(-m - \omega \Delta+ 2\omega) = 1\, ,
\ee
because of (\ref{reflexion formula for Gh}). Hence any pair of fields with a complex mass term is integrated for free in the localized partition function.

\bibliography{bibchiral}{}
\bibliographystyle{utphys}

\end{document}